\documentclass[12pt]{article}
\usepackage{amsmath,amsxtra,amssymb,latexsym,amscd,amsfonts,multicol,enumerate,ifthen,indentfirst,amsthm,amstext}
\usepackage[mathscr]{eucal}

\usepackage[utf8]{inputenc}
\usepackage{color}
\usepackage{graphicx}
\usepackage{epsfig}
\usepackage{float}
\usepackage{array,multicol, multirow , longtable}
\usepackage{pdflscape}
\usepackage[caption = false]{subfig}

\def \I{{\mathbf I}}
\def \1{{\mathbf 1}}
\def \E{{\mathbb E}}
\def \Pr{{\mathbb P}}
\def \P{{\mathcal P}}

\def \N{{\mathcal N}}
\def \Z{{\mathscr Z}}

\def \Er{{\text  {Er}}}

\def \Re{{\text  {Re}}}

\newtheorem{theorem}{Theorem}[section]
\newtheorem{lemma}[theorem]{Lemma}
\newtheorem{proposition}[theorem]{Proposition}

\newcommand*\colvec[3][]{
    \begin{pmatrix}\ifx\relax#1\relax\else#1\\\fi#2\\#3\end{pmatrix}
    }
\begin{document}


\begin{center}
{\Large
	{\sc  Accuracy of areal interpolation methods}\\{\sc for count data}
}
\bigskip

 DO Van Huyen $^{1}$ \& Christine THOMAS-AGNAN $^{2}$ \& Anne VANHEMS $^{3}$
\bigskip

{\it
$^{1}$ huyendvmath@gmail.com

$^{2}$ Christine.Thomas@tse-fr.eu

$^{3}$ a.vanhems@tbs-education.fr
}
\end{center}
\bigskip

{\bf Abstract.}
The combination of several socio-economic data bases originating from different administrative sources
collected on several different partitions of a geographic zone of interest into administrative units induces the so called areal interpolation problem. This problem is that of allocating  the data from a set of source spatial units to  a set of target spatial units. A particular case of that problem is the re-allocation to a single target partition which is a regular grid. At the European level for example, the EU directive 'INSPIRE', or INfrastructure for SPatial
InfoRmation, encourages the states to provide socio-economic data on a common grid to facilitate economic studies across states. In the literature, there are three main types of such techniques: proportional weighting schemes, smoothing techniques and
regression based interpolation. We propose a stochastic model based on Poisson point patterns to study the statistical accuracy of these techniques for regular grid targets in the case of count data. The error depends on the nature of the target variable and its correlation with the auxiliary variable.  For simplicity, we restrict attention to proportional weighting schemes and Poisson regression based methods. Our conclusion is that there is no technique which always dominates.

\smallskip



\section{Introduction}
The analysis of socio-economic data often involves the  integration of various spatial data sources. Those data are often independently collected  by a variety of offices and for different purposes. The zonal set systems used by distinct offices are rarely  compatible and this leads to many difficulties. The problem of merging data bases on different spatial supports is called the areal interpolation or basis change problem (Goodchild and Lam 1980). In France, the need for official statistics at a more and more refined territorial level has been recognized by INSEE. In Europe, one of the objectives of the EU directive ``INSPIRE'', for INfrastructure for SPatial InfoRmation, is to harmonize quality geographic information to support the formulation and evaluation of public policies and activities which directly or indirectly impact the environment. Many statistical methods are proposed in the literature to handle this problem (dasymetric methods, regression methods, smoothing techniques) and the reader is referred to Do et al. (2014) for a recent review of the simplest ones. The problem of their relative accuracy is most often treated at an empirical level (see for example Reibel and Bufalino (2005), Mennis (2006), Flowerdew and Green (1992), Flowerdew et al. (1991), Reibel and Agrawal (2007), Gregory (2002)). At the theoretical level, only few articles address this problem (Sadahiro (1999, 2000)) and this is the objective of this work.

  Comparing the accuracy of the different methods is difficult because the relative accuracy depends on several factors: nature of the target variable, correlation between the target and auxiliary variables, shapes of zonal sets, relative size  between the two zonal sets,... In order to derive theoretical results, we need to consider simplifying restrictions. For this reason, in this document, we first of all restrict attention  to data obtained from counts (see section \ref{count}): they are frequent in the literature and cover most of the  cases in the socio-economic applications. We also restrict the comparison to the simplest classes of methods which are the dasymetric and the regression ones. At last, we make the assumption that  target zones are nested within source zones. Indeed, this is  not really a restriction since the intersections between sources and targets are always nested within sources and it is immediate to go from intersection level to target level by aggregating the predictions as we will see later. In section \ref{count}, we define what we mean by data obtained from counts and we introduce a mathematical model adapted to this case. In order to illustrate the methods and check our theoretical results, we present  two sets of simulated data that we use later. In section \ref{methods}, we recall the formulas for the dasymetric and Poisson regression areal interpolation. Finally, in section \ref{acc1}, we compare the relative accuracy of areal weighting and dasymetric methods with finite distance results whereas in section \ref{acc2}, we compare the relative accuracy of  dasymetric and Poisson regression methods with
asymptotic methods. In both sections, we comment the results obtained on the toy examples presented in section \ref{count}. All proofs are in the appendix.
\section{Count data and Poisson point pattern model}
\label{count}

The variable of interest $Y$ that needs to be interpolated is called the {\bf target variable} and it needs to have a meaning on any subregion of the given space.  $Y_D$ will denote the value of the target variable on the subregion $D$ of the region of  interest  $\Omega.$

\noindent In the general area-to-area reallocation problem, the original data for the target variable  is available for a set of  source zones $S_s$  ($s= 1, \cdots, n_S,$) and has to be transferred to an independent set of target zones $T_t$ ($ t=1, \cdots, n_T$). The variable $Y_{S_s}$ will be denoted by $Y_s$ for simplicity and similarly for $Y_{T_t}$ by $Y_t$. The source zones and target zones are not necessarily nested and their boundaries do not usually coincide.

\noindent  Overlaps between the two sets are called intersection zones and denoted by $A_{st}$ for the intersection between the source $S_s$ and the target $T_t$. For simplicity,  $Y_{A_{st}}$ will be denoted by $Y_{st}.$ Many methods involve the areas of different subregions (sources, targets or other). We will denote by $\mid A \mid$ the area of any subregion $A$.

\noindent  Most of economic data collected at regional level result from aggregating point data and are only released in this aggregated form. Intuitively, let us say that a point data set is a set of a random number of random points in a given region of geographical space. The collection of corresponding numbers of such points in given subdivisions of this region is a count data set.
For example with census data, a population count on a given zone is the number of inhabitants of the zone. This number is obtained from the knowledge of the addresses of these people. The collection of coordinates of such addresses is the underlying point data set. Examples of areal interpolation of population or subpopulation counts can be found for example in
Goodchild and Lam (1980),  Langford (2005), Mennis and Hultgren (2006), Reibel and Agrawal (2007). Other types of counts are encountered frequently, for example number of housing units in Reibel and Bufalino (2005).
 Another frequent type of count related variable is the number of points per areal unit associated to a point data set: it is a density type variable. Examples of areal interpolation of population densities can be found in Yuan et al. (1997) and Murakami (2011). An even more general type is when the variable is a ratio of counts such as number of doctors per patient. There is an easy one to one correspondence between a count variable and a density variable which allows to transform one type into the other so that
 any treatment of counts can be extended to densities and reversely.
 A count variable belongs to the family of extensive variables, which are variables whose value on a region is obtained by summing up its values on any partition into subregions (aggregation formula hereafter). A density variable belongs to the family of intensive variables, which are variables whose value on a region is obtained from values on any partition into subregions by a weighted sum (see Do et al. 2014 for more details). In the case of population density, the weights are given by the areas of the subregions of the partition. In the remainder of this paper, we will concentrate on pure count variables.

 We introduce a model for an extensive count variable by assuming that there exists an underlying (unreleased) Poisson point pattern $Z_Y$ (in the population example, the positions of the individuals of the population) and that the target variable $Y$ on a subzone $A$ is the number of points of $Z_Y$ in $A$. For a partition $\Omega_i, i=1,2,...,k$ of the region $\Omega$, the extensive property is clearly satisfied

$$Y_\Omega = \sum_{i=1}^k Y_{\Omega_i}$$

\noindent  With the proposed Poisson point pattern assumption, for any zone $A$, $Y_A= \sum_i \1_A(Z_i)$ is a Poisson distributed random variable with mean $\lambda_A =\int_A \lambda_{Z_Y} (s) ds,$ where $\lambda_{Z_Y}$ is the intensity of the point process $Z_Y$.

\noindent This model implies that $Y_A$ and $Y_B$ are automatically independent for all disjoint couples of subregions $A$ and
$B$ due to the Poisson process nature. We could use point pattern models with interaction effects while retaining the extensive property but we rather devote this article to this first case, keeping the interaction case for further developments.

\noindent As we will see in the next section, some methods we want to compare (dasymetric and univariate regression) make use of an auxiliary information. For the auxiliary variable $X$ to be relevant, there must be some relationship between the target variable and the auxiliary variable. In many cases a categorical information is used such as land cover: Reibel and Agrawal (2007) and Yuan et al. (1997) use land cover type data on a 30 meters resolution grid,
Mennis and Hulgren (2006) use 5 types of land cover obtained manually from aerial photography.
Li et al. (2007) just use a binary information such as  unpopulated versus populated zones.
 Reibel and Bufalino (2005) interpolate the 1990 census tract counts of people and housing using length of streets as auxiliary information. Mugglin and Carlin (1998) exploit population to interpolate the number of leukemia cases.
 The use of a  continuous auxiliary information can also be found: Murakami (2011) utilizes distance and land price to predict population density. In the rest of the paper, we concentrate on a single extensive auxiliary variable $X$ that is also a count in order to be able to consider the accuracy of all methods simultaneously (more details at the end of section \ref{methods}). Therefore it corresponds to another underlying point process $Z_X$ with intensity $\lambda_{Z_X}$.

\noindent The auxiliary variable $X,$  has to be known at intersection level in the case of dasymetric and at the target level in the case of regression. We need to write a formal relationship between our target variable and the auxiliary information.  The model we propose assumes that the following relationship holds between the two underlying point processes intensity functions
\begin{equation}\label{intens}
\lambda_{Z_Y} (s) = \alpha + \beta \lambda_{Z_X}(s),
\end{equation}
\noindent where $s$ is location.
\noindent Therefore, the following relationship holds between  $Y$ given $X$:
at the level of any subset $A$ of the region, the conditional distribution of $Y_A$ given $X_A = x_A$ is given by

\begin{equation}\label{model}
Y_{A} \sim \P(\alpha |A|+ \beta x_A)
\end{equation}

\noindent This relationship will be  used at target level $A = T$ and at source level $A=S$. This model in its general form will be called auxiliary information model (AIM). In this model, the intensity of $Z_Y$ is driven by two effects: the effect of the auxiliary variable $X$ and the effect of the area of the zone.  If we look at target level, the target variable is Poisson distributed with a mean comprising two parts $\E(Y_T)=\alpha|T| +\beta x_T$: the first part $\alpha |T|$ reflects the impact of the area of the zone $T$, whereas $\beta x_T$ is the impact of the auxiliary variable.  The linearity of the expected value of $Y$  with respect to the area and to the auxiliary information is not canonical in a Poisson regression model for counts but in our case it derives naturally from \eqref{intens}.

\noindent In sections \ref{homAIM} and \ref{phomAIM}, we introduce two sub-models of model \eqref{model} depending on the intensity function $\lambda_{Z_Y}$.  We consider the case of a constant intensity (homogeneous model) and the case of a piecewise constant intensity (piecewise homogeneous model).
\section{Methods}
\label{methods}
\subsection{Prediction techniques}

Do et al.(2014) classify areal interpolation methods into three groups: smoothing, dasymetric and regression based
methods. We discard  smoothing since it is concerned with continuous target variables which is not adapted to our count data model.
We therefore focus here on the remaining two groups: dasymetric and regression based methods.

Dasymetric is a class of methods using a weighting scheme to allocate the original data to the intersections and then applying an aggregation step to get to target level. The simplest method in the dasymetric class is the areal weighting interpolation which uses area as weighting scheme:  the data is allocated to the targets based on the assumption that the target variable is homogeneous at source level:

$$\hat{Y}_t=\sum_{s: s \cap t \not= \emptyset} \hat{Y}_{st}=\sum_{s: s \cap t \not= \emptyset} \dfrac{|A_{st}|}{|S_s|} {Y}_{s}.$$

\noindent  Note that areal interpolation does
not use any auxiliary information other than area which is usually available.

The general dasymetric method is supposed to improve upon the areal weighting interpolation method when an additional variable is known to be linked to the target variable leading to alternative weighting schemes.
Voss et al. (1999) use road segment length and the number of road nodes for allocating demographic characteristics. Population, which is collected at fine levels in general, is often used as an auxiliary information for other variables like in Gregory (2002) or Mugglin and Carlin (1998).
Instead of homogeneity, the dasymetric method with auxiliary information $X$ assumes that the target variable is proportional to the auxiliary variable at intersection level.

$$\hat{Y}_t=\sum_{s: s \cap t \not= \emptyset} \hat{Y}_{st}=\sum_{s: s \cap t \not= \emptyset} \dfrac{X_{st}}{X_s} {Y}_{s}$$

\noindent where $X_s=\sum_t X_{st}$. This entails that $X$ has to be known at intersection level, which is quite restrictive.

 Concerning the regression based methods, there are several types of regression based methods also involving auxiliary information (see Do et al, 2014). Given the nature of the target variable in our model (\ref{model}), we concentrate on the Poisson regression presented in Flowerdew et al. (1991) for the purpose of predicting population (which is an extensive variable) with categorical auxiliary information. Based on model (\ref{model}), a Poisson regression with identity link is performed at source level yielding estimators $\hat \alpha, \hat \beta$ for the parameters $\alpha$ and  $\beta.$

\noindent  The prediction of the target variable at intersection level is then obtained by
\begin{equation}\label{RegPredictor}
\hat{Y}^{REG}_{st}=\hat{\alpha} |A_{st}|+ \hat{\beta} X_{st}
\end{equation}

\noindent   and the final step aggregates intersections predictions at target levels. The regression based methods can be considered as more powerful than the dasymetric methods in the sense that they can incorporate multivariate auxiliary information and that the knowledge of auxiliary information is only needed at target level and not at intersection level. However, the purpose of this paper being to compare the accuracy of dasymetric methods and Poisson regression methods from a methodological point of view and for the case of extensive count data, we therefore concentrate on the unidimensional auxiliary count variable case.

\noindent One property often quoted concerning these methods is the pycnophylactic property. This property requires the preservation of the initial data in the following sense at some geographical level: at source level for example, it means that the predicted value for source $S_s$ obtained by aggregating the predicted values on intersections with $S_s$ should coincide with the observed value on $S_s$. The enforcement of this property will allow us to introduce an improved version of the basic Poisson regression method.

\subsection{Prediction error criteria}
The accuracy assessment necessitates the choice of a prediction  error criterion and of a geographic level. In this framework, examples of  criteria are root mean square error or mean square error (Sadahiro 1999, Reibel 2006,...) at regional level (that is the union of all sources), or relative absolute error at target level (Langford, 2007).
We denote by MET a generic method of prediction and let $MET$ be $DAW$ for the areal weighting method, $DAX$ for the general dasymetric method, $REG$ for the Poisson regression method and $ScR$ for the scaled regression method which will be presented later in section \ref{acc2}. We recall that we assume all target zones are nested within  source zones.

\noindent In section \ref{acc1}, we use mean square error at source level to compare the areal weighting and dasymetric methods.  For method MET, the source level error is then computed as follows

\begin{equation}
\Er^{MET}_S=\sum_{t \subset S}\Er^{MET}_t=\sum_{t \subset S} \E(\hat{Y}^{MET}_t-Y_t)^2
\end{equation}
\noindent and the overall regional error is
\begin{equation}\label{overall}
\Er^{MET}=\sum_{S} \sum_{t \in S} \E(\hat{Y}^{MET}_t -Y_t)^2
\end{equation}

In section \ref{acc2}, we use mean square error at target level
\begin{equation}
\Er^{MET}_t=\E(\hat{Y}^{MET}_{t}-Y_t)^2
\end{equation}

\noindent to compare the dasymetric and Poisson regression methods.

\noindent In general, we will also use the relative error criterion defined as

\begin{equation}
\Re^{MET}_S={\sqrt{\Er^{MET}_S} \over \E(Y_S)}
\end{equation}
where $\Re^{MET}_S$ is the relative error of method $MET$ at source level for source $S$ with method $MET$.

\section{Relative accuracy of areal weighting and dasymetric: finite distance assessment}
\label{acc1}

Let us briefly summarize the findings of the assessments found in the literature for the comparison of general dasymetric and areal weighting.
 For empirical assessments, several authors report that the dasymetric method improves upon areal weighting. Depending on the context, the improvement varies: Langford (2007) reports improvements of 54\%, 57\%, and 59\% better depending on the auxiliary information used; Reibel and Bufalino (2005) reports improvements of 71.26\% and 20.08\% with street length auxiliary information for the two target variables: housing units and total population. For theoretical assessments, Sadahiro (1999, 2000) compares the areal weighting interpolation and the point-in-polygon method with a theoretical model. We did not mention yet the point-in-polygon method because it is a very elementary one consisting in allocating a source value to the target which contains its centroid. Using a stochastic model, he finds that the factors that impact the accuracy of the methods are the size and shape of target and source zones, the properties of underlying points.

\noindent In this section, we prove some theoretical properties in subsection \ref{genAIM}, with two particular cases in \ref{homAIM} and \ref{phomAIM},  and a toy example in \ref{firsttoy}.

\noindent Since targets are nested within sources, the predictors of the two methods depend only on the source that contains the concerned target zone. For that reason, we focus on studying one source zone denoted by $S$. For a target $T$ in $S$, the two predictors are as follows
\begin{equation}\label{DAW}
\hat{Y}^{DAW}_{T}=\dfrac{|T|}{|S|}Y_S
\end{equation}
and
\begin{equation}\label{DAX}
\hat{Y}^{DAX}_{T}=\dfrac{x_T}{x_S}Y_S
\end{equation}

\subsection{General auxiliary information model}
\label{genAIM}
Lemma \ref{bias} gives the expression of the prediction bias and variance in model AIM for areal weighting interpolation and dasymetric interpolation at target level.
 \begin{lemma}\label{bias} In model AIM, the prediction biases and variances
 of areal weighting interpolation and dasymetric methods are given by
  \begin{align}
  \E(\hat{Y}^{DAW}_{T}-Y_T)&=\beta x_S (\dfrac{|T|}{|S|}-\dfrac{x_T}{x_S})\label{BiasTW}\\
  \E(\hat{Y}^{DAX}_{T}-Y_T)&= \alpha |S| (\dfrac{x_T}{x_S}-\dfrac{|T|}{|S|})\label{BiasTx}\\
    Var(\hat{Y}^{DAW}_{T}-Y_T)&=\beta x_S (\dfrac{|T|}{|S|}-\dfrac{x_T}{x_S})^2+\beta x_T(1-\dfrac{x_T}{x_S}) +\alpha |T| (1-\dfrac{|T|}{|S|})\label{VarTW}\\
   Var(\hat{Y}^{DAX}_{T}-Y_T)&=\alpha |S| (\dfrac{|T|}{|S|}-\dfrac{x_T}{x_S})^2+\beta x_T(1-\dfrac{x_T}{x_S}) +\alpha |T| (1-\dfrac{|T|}{|S|})\label{VarTx}
    \end{align}
 \end{lemma}
 \noindent First note that the two biases have opposite signs, in other words, if the areal weighting interpolation method underestimates then the dasymetric method overestimates and vice versa. This fact can be interpreted as follows: while the true intensity comprises two effects, these methods treat only one of them which causes the contrast. Although the  signs of biases are opposite, their absolute values are both proportional to $\dfrac{|T|}{|S|}-\dfrac{x_T}{x_S}$ which measures the divergence between the share of the auxiliary information in target $T$ with respect to $S$ and the share of the area of $T$ with respect to $S$. This divergence is also proportional  to $\dfrac{x_S}{|S|}-\dfrac{x_T}{|T|}$ and hence can be viewed as a distance to proportionality between area and auxiliary information. The bias of the areal interpolation method with its assumption of homogeneity is independent in the areal effect $\alpha |S|$ but is proportional to the ignored auxiliary information effect, and reversely the dasymetric method which focuses on the effect of the auxiliary information gets rid of the $\beta x_S$ in its bias but is proportional to the ignored areal effect. We will build on this to propose a new method in the next section.

\noindent
The two variances have a common part  $\beta x_T(1-\dfrac{x_T}{x_S}) +\alpha |T| (1-\dfrac{|T|}{|S|})$ which we can interpret as the loss of information when transferring data from a large source zone to a smaller target zone. For the remaining part,  the same explanations as for the bias stands. Both variances have a parabola shape with respect to $x_T$ (respectively to $|T|$) with a maximum at $x_T=\dfrac{1}{2} x_S,$ (resp. $|T|=\dfrac{1}{2}|S|$): we can say loosely that the variances are maximum when the target zone is around a haft of the source. They vanish when the target zone is either  empty or coincide with the source which makes sense. The reallocation to a larger target intuitively decreases the difficulty of the disaggregation problem except that the error also depends on the expected number of points so we should turn attention to relative error. If one divides the variances by the square of the expected number of points in the target zone ${\mathbb E}(Y_T)$, we can see that the relative error will tend to zero as ${\mathbb E}(Y_T)$ tends to infinity.

\noindent Since the dasymetric method is pycnophylactic, the bias at source level is zero. Lemma \ref{variances} reports the expression of the prediction variances in model AIM for areal weighting interpolation and dasymetric interpolation at source level.

\begin{lemma}\label{variances} In model AIM, the variances of areal weighting and dasymetric methods at the source level are
    \begin{align}
    Var_S^{DAW}&=\beta x_S \sum_T(\dfrac{|T|}{|S|}-\dfrac{x_T}{x_S})^2+\beta x_S(1-\sum_T \dfrac{x_T^2}{x_S^2}) +\alpha |S| (1-\sum_T \dfrac{|T|^2}{|S|^2})\label{VarSW}\\
    Var_S^{DAX}  &=\alpha |S| \sum_T(\dfrac{|T|}{|S|}-\dfrac{x_T}{x_S})^2+\beta x_S(1-\sum_T \dfrac{x_T^2}{x_S^2}) +\alpha |S| (1-\sum_T \dfrac{|T|^2}{|S|^2}) \label{VarSx}
    \end{align}
    \end{lemma}

\noindent To get an insight at impact of the number $n_T$ of the target zones, we consider the special case where all  targets have the same size. In this case, $\dfrac{|T|}{|S|}=\dfrac{1}{n_T}$ for any $T$, and we get

 \begin{align*}
 Var_S^{DAW}&=(1 - \dfrac{1}{n_T})(\alpha |S|+\beta x_S )\\
  Var_S^{DAX}  &=(1 - \dfrac{1}{n_T})(\alpha |S|+\beta x_S  )+ \sum_T (\dfrac{x_T^2}{x_S^2} - \dfrac{1}{n_T})(\alpha |S|-\beta x_S )
   \end{align*}

It is obvious that the larger the number of the target zones, the larger the variances, which  agrees with our conclusion concerning the size of targets.  Indeed, when the area of the target zones gets smaller, the error on each target decreases but the total error at the source level gets larger due to the effect of the number of the targets.

\noindent We are now ready to compute the mean square error difference between  the two methods.
We introduce the following quantities which quantify a relative contribution of the corresponding effect to the overall mean at the geographical level of a subregion $A$: $$I_A(X)= \frac{\beta x_A}{\alpha |A| + \beta x_A}$$
$I_A(X)$ is the relative contribution of variable $X$ and similarly $I_A(|.|)= \frac{\alpha |A|}{\alpha |A| + \beta x_A}$ is the relative contribution of the areal effect.

The imbalance between the two effects is measured by the difference
$$\Delta_A = I_A(|.|) - I_A(X) = \frac{\alpha |A| - \beta x_A}{{\mathbb E}(Y_A)}.$$
This quantity ranges between $-1$ when there is a pure $X$ effect and $1$ when there is a pure areal effect with a value of zero when the two effects are of equal size.

\noindent We can derive from lemmas \ref{bias} and \ref{variances} the expression of the absolute and relative errors of the two methods at source level as a function of the relative contributions terms.

\begin{theorem}\label{reler}
\begin{align}
\Er^{DAW}_S=&I_S(X)^2 \E(Y_S)^2 D+ I_S(X) \E(Y_S)(D+B)+I_S(|.|)\E(Y_S)C \label{ErDAW}\\
\Rightarrow (\Re^{DAW}_S)^2&=I_S(X)^2 D+{1 \over \E(Y_S)}[ I_S(X)(D+B-C) +C]\label{ReDAW}
\end{align}
\begin{align}
(\Re^{DAX}_S)^2=I_S(|.|)^2 D +{1 \over \E(Y_S)}[I_S(|.|)(D-B+C) +B]\label{ReDAX}
\end{align}

where $D=\sum_T(\dfrac{|T|}{|S|}-\dfrac{x_T}{x_S})^2, B=1- \sum_T\dfrac{x_T^2}{x_S^2}, C=1- \sum_T\dfrac{|T|^2}{|S|^2}$ are positive.
\end{theorem}

\noindent Note that $B,C$ and $D$ only depend on the the geometry of the problem and the auxiliary information, whereas the relative contribution terms and $\E(Y_S)$ depend on the coefficients $\alpha$ and $\beta$. It is interesting to mention the symmetry between the two methods which stands clearly in these formulas when we exchange the two contributions terms. One can derive from this theorem the difference between the relative errors of the two methods

\begin{equation}\label{DiffReDAWDAX}
(\Re^{DAW}_S)^2-(\Re^{DAX}_S)^2=-D * \Delta_S (1+{1 \over \E(Y_S)})
\end{equation}

\noindent which turns out to be clearly proportional to the imbalance term $\Delta_S$.  Similarly, one can approximate the ratio of the two relative errors when $\E(Y_S)$ is large on the target $A=T$ and on the source $A=S$ by

\begin{align}
 {\Re^{DAW}_A\over \Re^{DAX}_A}&\approx {I_A(X) \over I_A(|.|)}\label{RaReDAWDAX}
\end{align}

\noindent This ratio roughly ranges from $0$ to $+\infty$ at the extreme cases of a pure $X$ or areal effect showing that one can outperform the other by a large amount.
\noindent Let us now turn attention to the difference between the two errors.

 \begin{theorem} \label{TheoDiffErDaxDaw} The difference between the errors of areal weighting and dasymetric methods on a target zone $T$ is
  \begin{align*}
  \Er^{DAW}_T&-\Er^{DAX}_T=(\dfrac{|T|}{|S|}-\dfrac{x_T}{x_S})^2\Delta_S {\mathbb E}(Y_S)({\mathbb E}(Y_S)+1)
  \end{align*}

  \end{theorem}

  \noindent The important conclusion of this result is that the sign of the difference in error agrees with the sign of $\Delta_S$, i.e. the sign of $(\alpha |S| - \beta x_S)$. Moreover, the stronger the effect of the auxiliary information $I_S(X)$ is, the better the dasymetric method and the larger the difference between the two methods.

  This computation result leads to a very interesting consequence: if one of two effect dominates on a given source, the related  method wins on all target zones belonging to the source. It also shows that two methods will have the same accurracy if the two effects are balanced or the auxiliary variable is homogeneous.

  The normalized difference between the two effects $\Delta_S$ clearly determines which method is best.

\noindent At this point, it seems natural to look for a linear combination of these two predictors

\begin{align}\label{compositeclass}
\hat{Y}^C_T (w)&= w\hat{Y}^{DAW}_{T}+(1-w)\hat{Y}^{DAX}_{T},
\end{align}

\noindent which would combine their good properties. It turns out that in the class of linear combinations of areal weighting and dasymetric predictors, the best predictor is given by the following theorem

\begin{theorem}\label{composite}
In model AIM, the best predictor in the sense of minimizing (with respect to the weight $w$) the errors on any target zone $T$ in the class \eqref{compositeclass} is
\begin{equation}\label{Com}
\hat{Y}^C_T =\hat{Y}^C_T (w^*)=\dfrac{\alpha |T|+\beta x_T}{\alpha |S|+\beta x_S}Y_S
\end{equation}
for $w^*=\dfrac{\alpha |T|}{\alpha |S|+\beta x_S}$. Its error and relative error are respectively given by
\begin{align}
\Er^C_T&=  \dfrac{\lambda_T(\lambda_S - \lambda_T)}{\lambda_S} \label{ErComDAX}\\
(\Re^C_S)^2&=\dfrac{1}{4\E(Y_S)}[\Delta_S^2 D+2\Delta_S (C-B)+D+2B+2C]\label{ErComDAW}
\end{align}
Moreover, this predictor coincides with  the best linear unbiased predictor in model AIM.
\end{theorem}

\noindent Because $\dfrac{\lambda_T(\lambda_S - \lambda_T)}{\lambda_S} = Var(\hat{Y}^{DAX}_T-Y_T)-\lambda_S(\dfrac{x_T}{x_S}-\dfrac{\lambda_T}{\lambda_S})^2
= Var(\hat{Y}^{DAW}_T-Y_T)-\lambda_S(\dfrac{|T|}{|S|}-\dfrac{\lambda_T}{\lambda_S})^2 ,$ the prediction error of the  best predictor is smaller than the variances of the other two methods and the distance is the more important that the auxiliary information is further from homogeneity. Of course, $\hat{Y}^C_T$  is not a feasible predictor since it depends on the unknown coefficients $\alpha$ and $\beta$ of model AIM but we will use it as a benchmark tool on the one hand and we will relate it later on to the regression predictor. If we look at the error at the level of source $S$, we have that
$\Er^C_S= \lambda_S - \sum_{T}\dfrac{\lambda_T^2}{\lambda_S}\leq \lambda_S-\dfrac{\lambda_S}{n_T(S)},$ where $n_T(S)$ is the number of targets in source $S$, and hence this predictor's accuracy is worse when all targets have the same expected number of points $\dfrac{\lambda_S}{n_T(S)}$. It is interesting to note that the relative error (at source level $S$) of the best predictor tends to zero as the expected number of points in the source $S$ tends to infinity, which was not the case for the dasymetric methods. For a fixed expected number of points in a given source $S$, we can easily find the value of the imbalance $\Delta_S$ which minimizes the relative error of $\hat{Y}^C_T$ $\Delta^* = \dfrac{B-C}{D}=\dfrac{\sum_T (\dfrac{|T|}{|S|})^2-(\dfrac{x_T}{x_S})^2}{\sum_T (\dfrac{|T|}{|S|}-\dfrac{x_T}{x_S})^2}$ and thus derive a lower bound for the relative error for a given geometry.

\noindent Because intuitively, it is natural to think that areal weighting should be outperformed by dasymetric when the underlying process is inhomogeneous, we consider the two cases of homogeneous and piecewise homogeneous submodels.

\subsection{Homogeneous model}
\label{homAIM}
Areal weighting  interpolation is a simple and natural rule which is based on the assumption that the target variable is homogeneous at source level. Indeed in model AIM, it is equivalent to assume that the point process is homogeneous and its intensity is therefore  constant (equal to $\alpha > 0$) leading to:
$$Y_A \sim \P(\alpha |A|).$$
Substituting  $\beta=0$ in \eqref{BiasTW}, \eqref{VarTW}, \eqref{VarSW} we get the bias, variance and error in this case:
\begin{align*}
\E(\hat{Y}^{DAW}_{T}-Y_T)&=0\\
\Er^{DAW}_T&= Var(\hat{Y}^{DAW}_{T}-Y_T)=\alpha |T| (1-\dfrac{|T|}{|S|})\\
\Er^{DAW}_S&= Var_S^{DAW}=\alpha |S| (1-\sum_T \dfrac{|T|^2}{|S|^2})
\end{align*}

 Since $\dfrac{1}{n_T} \leq \sum_T \dfrac{|T|^2}{|S|^2} \leq 1$, the error at source level is maximum when all target zones have the same size, and minimal when there is a unique target which coincides with the source.

Substituting $\beta=0$ in \eqref{Com} leads to the conclusion that the best linear unbiased predictor in the homogeneous AIM model is given by the areal weighting method which is a natural result.
Let us now turn attention to a very simple non homogeneous model to illustrate the intuitive fact that the areal weighting interpolation method is not the best choice in a non homogeneous situation.
\subsection{Piecewise homogeneous model}
\label{phomAIM}
 Suppose the source zone $S$ comprises two homogeneous subzones $C_1$ and  $C_2$ called control zones with intensities $\alpha_1$ and $\alpha_2$ respectively. In this case, we get
 $$Y_A\sim \P(\alpha_* |A|)$$
where $A \subset C_*$ with $*=1,2$.
 For simplification reasons, we assume the target zones to be nested within the control zones. The results of lemmas \ref{bias} and \ref{variances} give in this case

\begin{align*}
&\E(\hat{Y}^{DAW}_{T}-Y_T)_{T: T \subset C_1}=\dfrac{|T|}{|S|}(\alpha_2-\alpha_1) |C_2|\\
&\E(\hat{Y}^{DAW}_{T}-Y_T)_{T: T \subset C_2}=\dfrac{|T|}{|S|}(\alpha_1-\alpha_2) |C_1|\\
 &Var_S^{DAW}=\alpha_1|C_1| (1-\sum_{T: T\subset C_1} \dfrac{|T|^2}{|S|^2})+\alpha_2|C_2| (1-\sum_{T: T\subset C_2} \dfrac{|T|^2}{|S|^2})\\
 &\Er^{DAW}_S=Var_S^{DAW}+\sum_{T: T\subset C_1}\dfrac{|T|^2}{|S|^2}(\alpha_2-\alpha_1)^2 |C_2|^2+\sum_{T: T\subset C_2}\dfrac{|T|^2}{|S|^2}(\alpha_2-\alpha_1)^2 |C_2|^2
\end{align*}

The variance has a similar structure to the one of the homogeneous model. The bias clearly shows that the difference between the two intensities of the subzones will drive the size of the error.

\subsection{Toy example}
\label{firsttoy}
\begin{figure}[h!]
\begin{center}
\subfloat[$Y_1$ at cells]{\includegraphics[width = 1.5in]{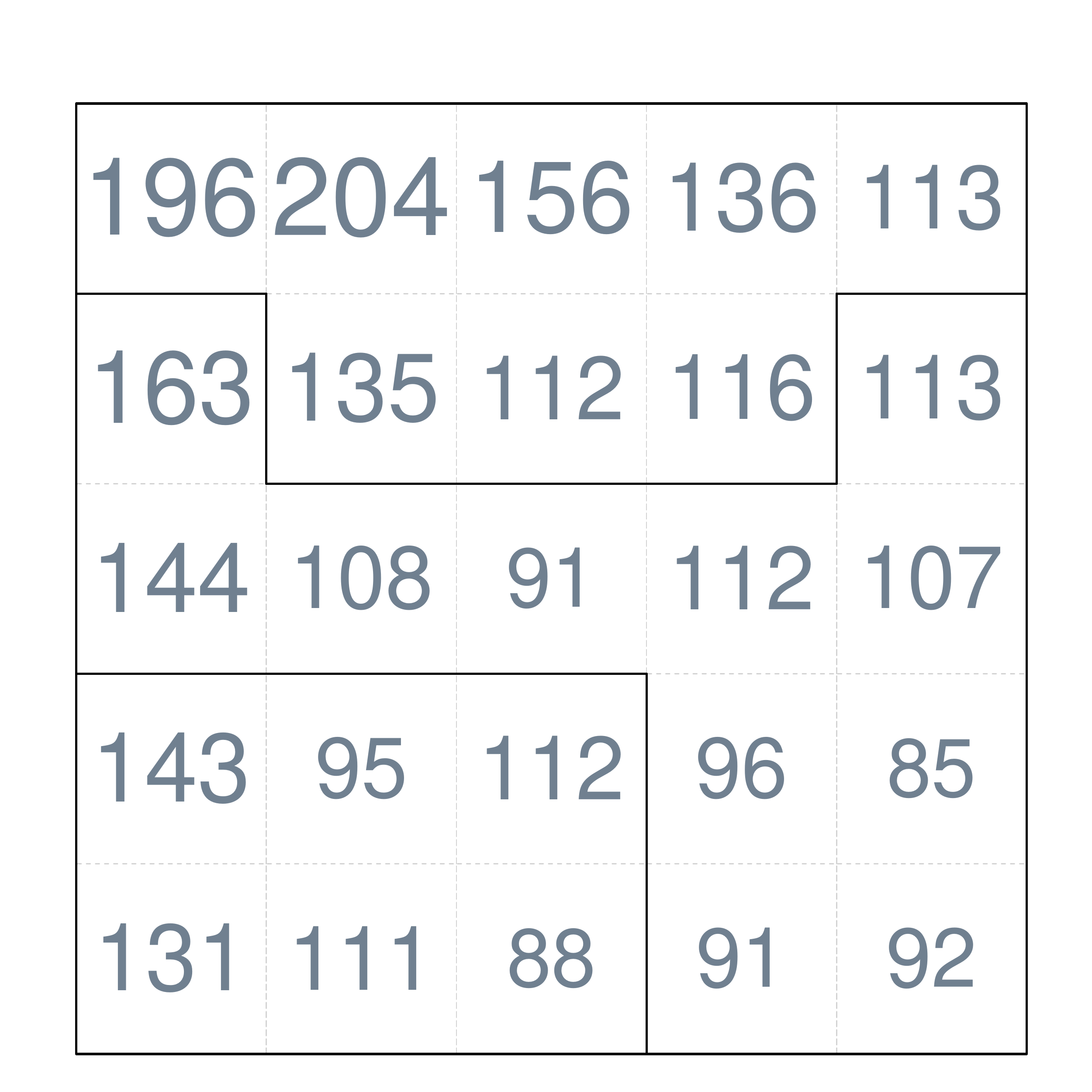}}~
\subfloat[$X$ at cells]{\includegraphics[width = 1.5in]{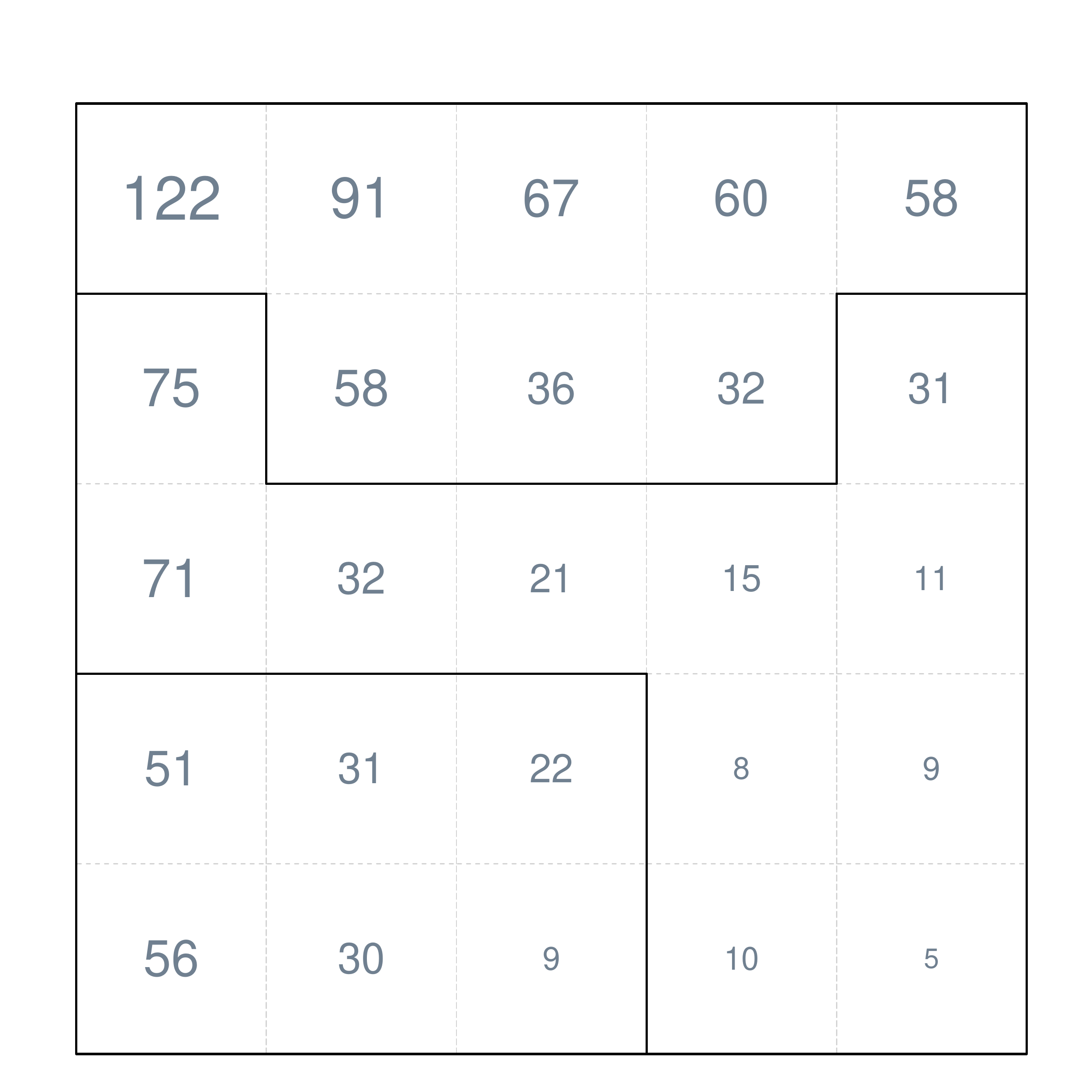}}~
\subfloat[$Y_2$ at cells]{\includegraphics[width = 1.5in]{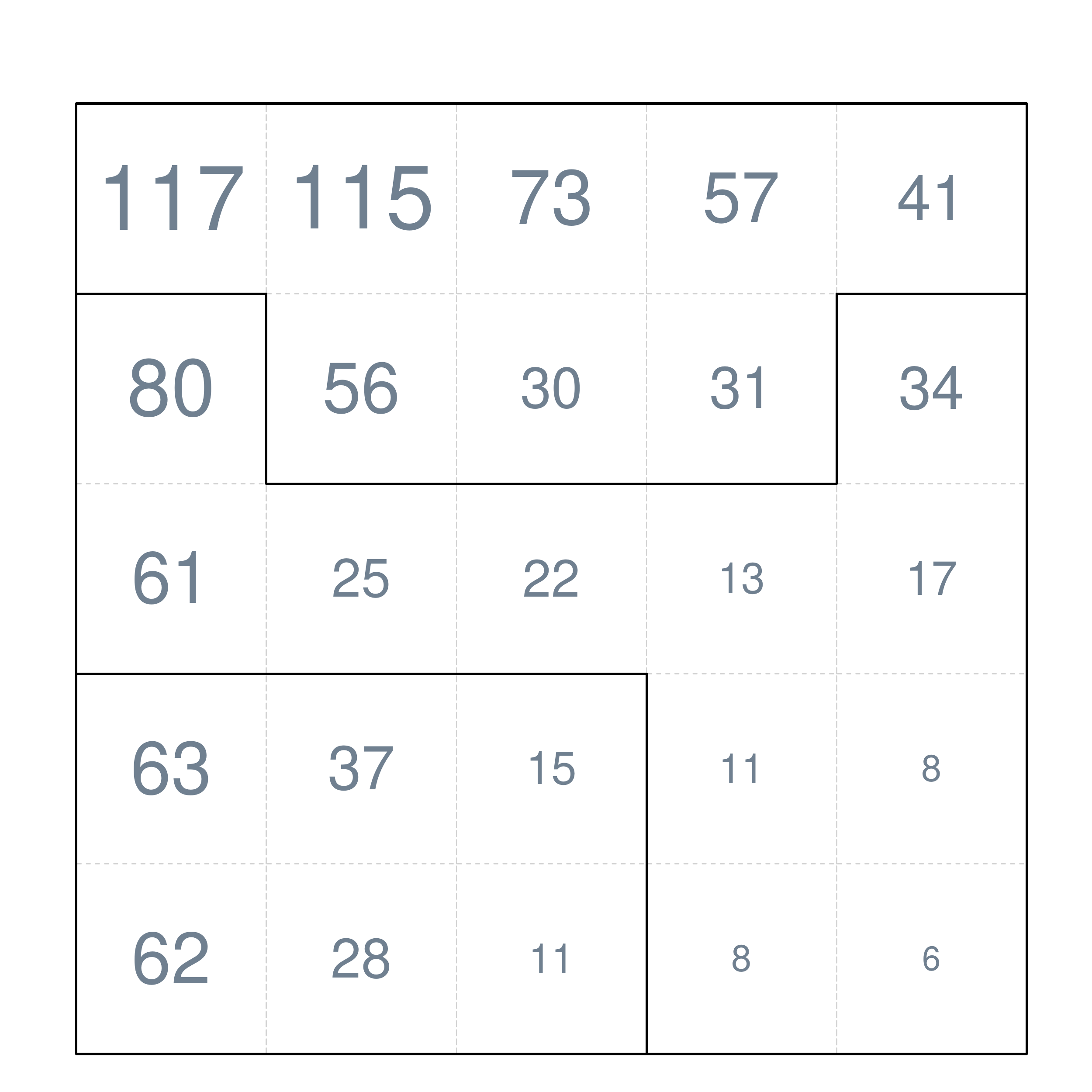}}~
\subfloat[Targets]{\includegraphics[width = 1.5in]{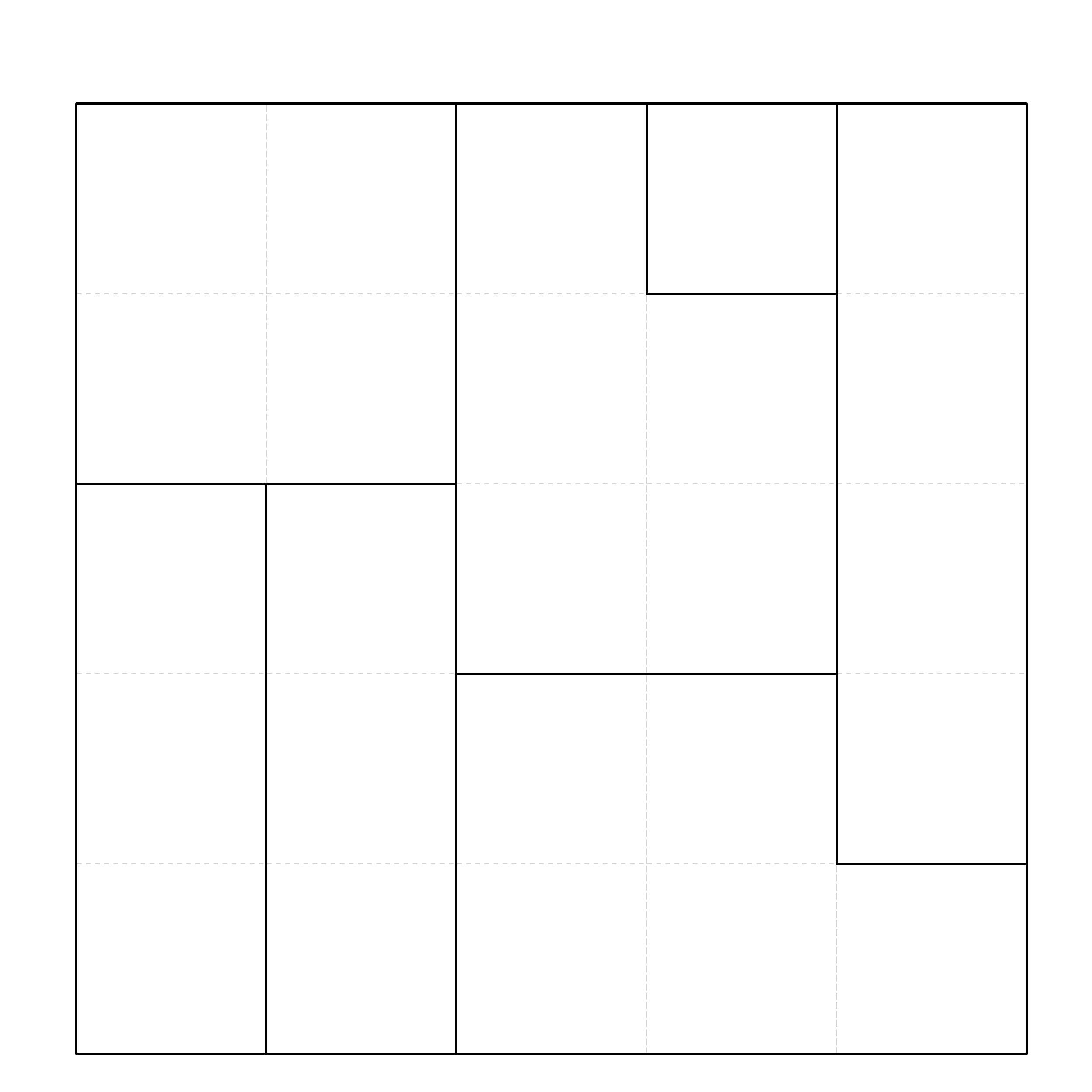}}~\\
\subfloat[$Y_1, Y_2$ at sources]{\includegraphics[width = 1.5in]{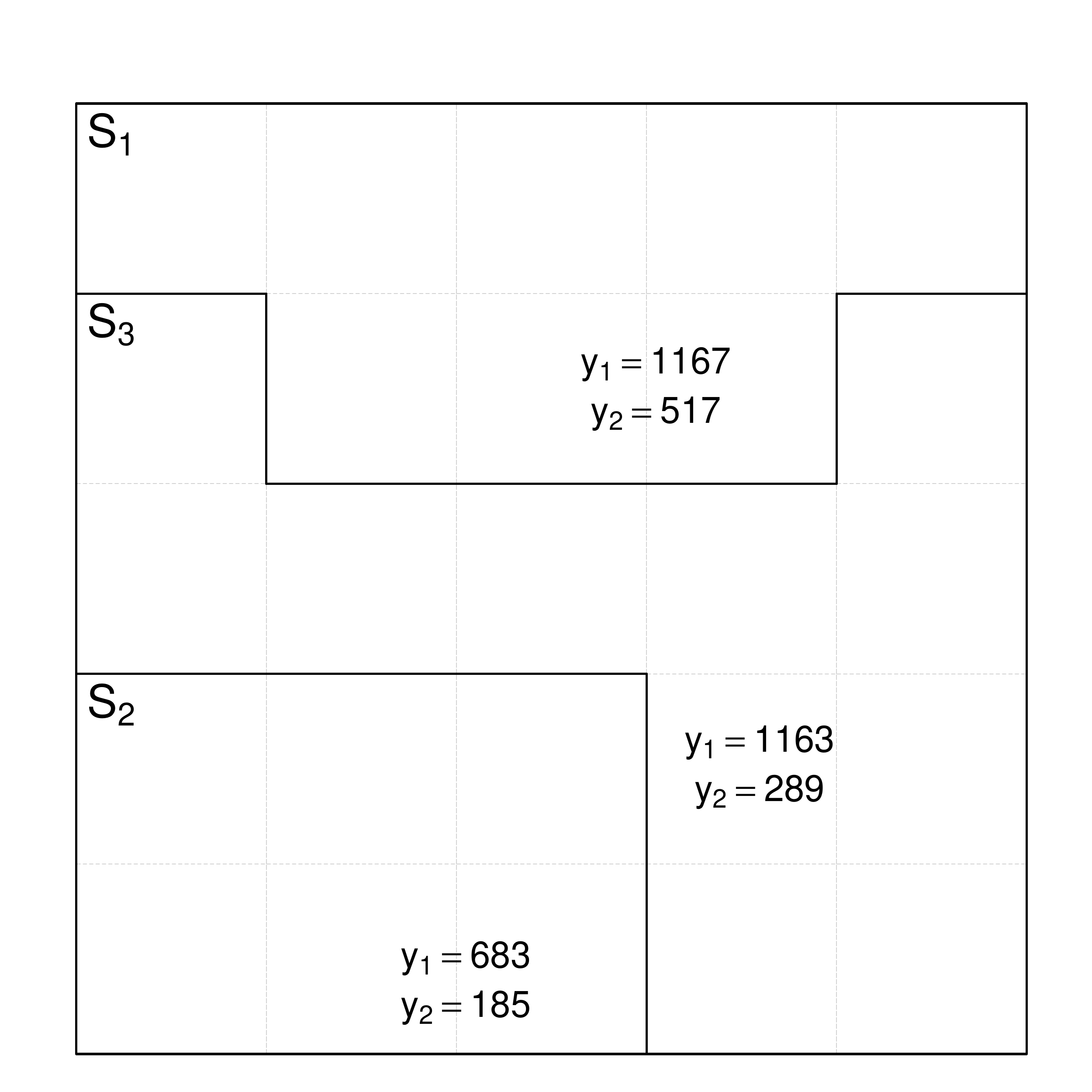}}~
\subfloat[$X$ at intersections]{\includegraphics[width = 1.5in]{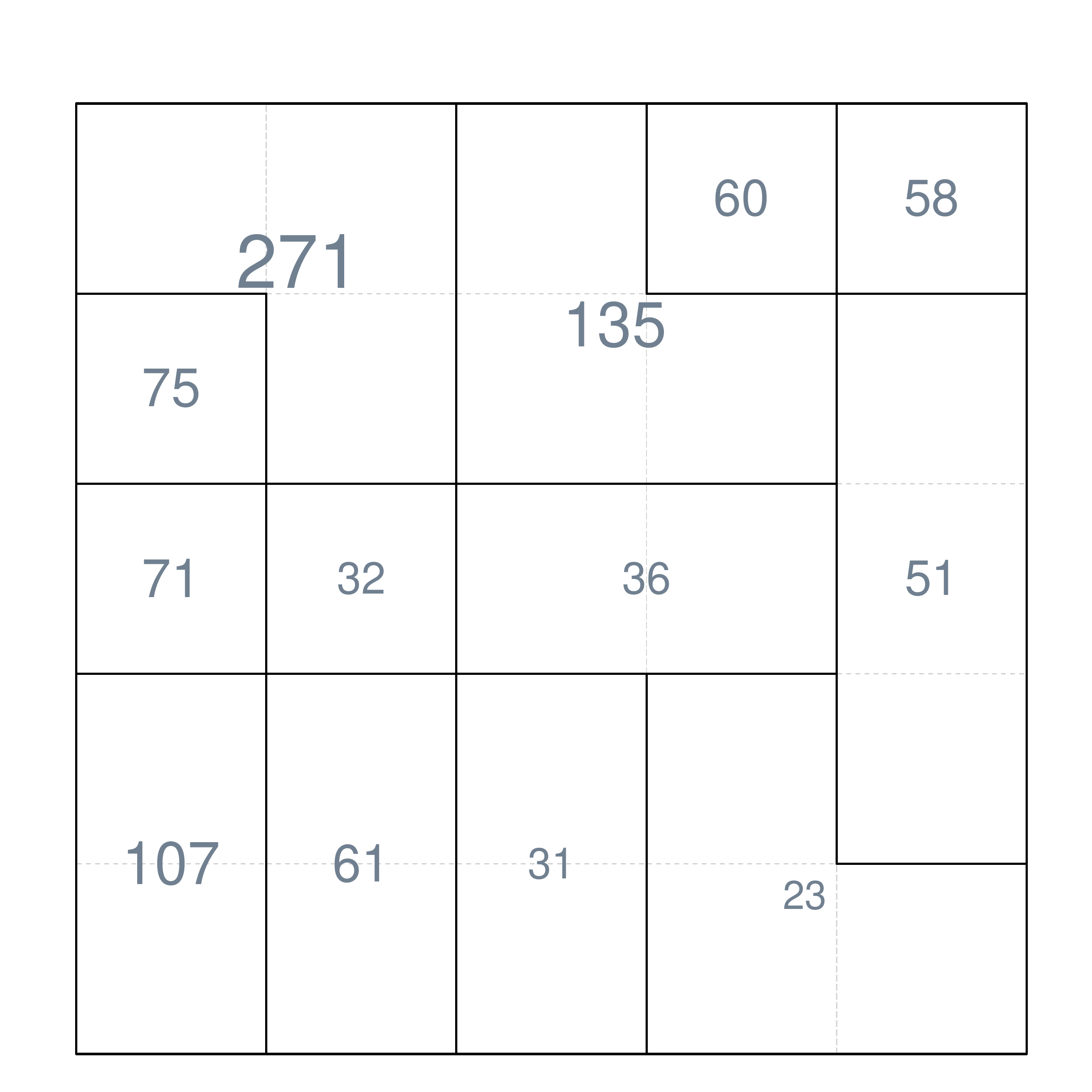}}~
\subfloat[$Y_1, Y_2$ at targets]{\includegraphics[width = 1.5in]{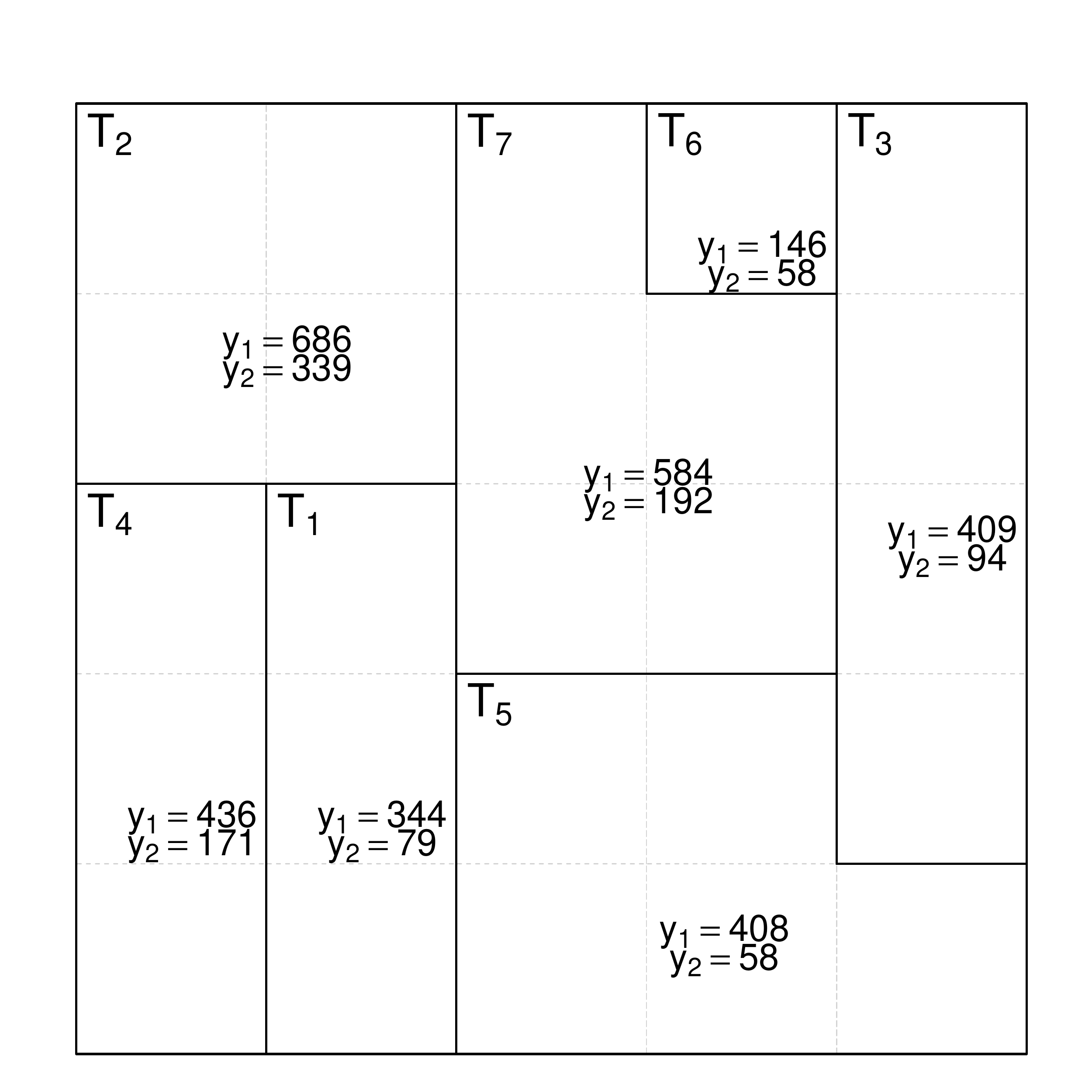}}
\end{center}
\caption{Toy example: Data at source, target, intersection zones.}\label{toyex1}
\end{figure}

\begin{figure}[h!]
\begin{center}
\subfloat[DAW for $Y_1$]{\includegraphics[width = 1.5in]{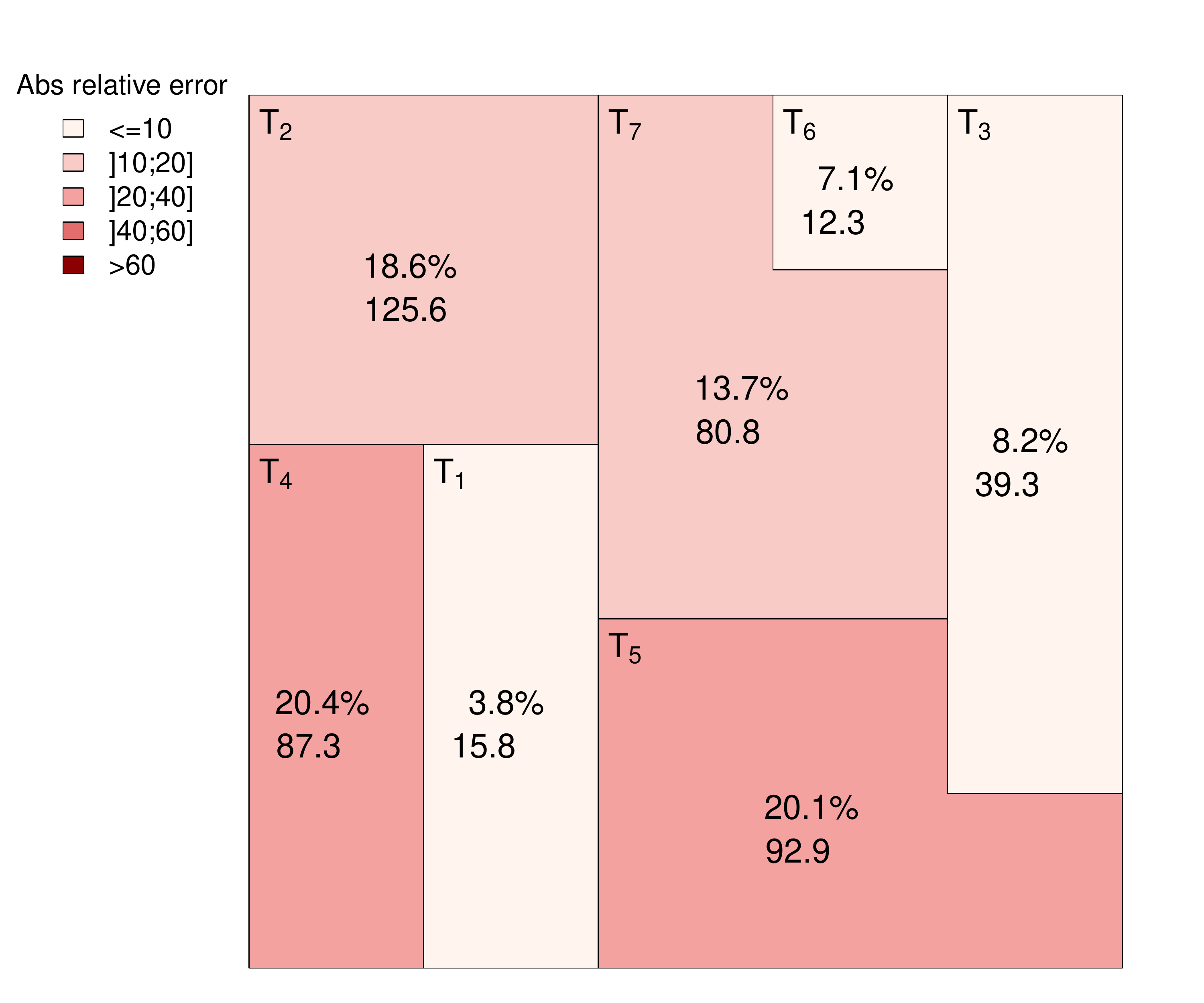}}
\subfloat[DAX for $Y_1$]{\includegraphics[width = 1.5in]{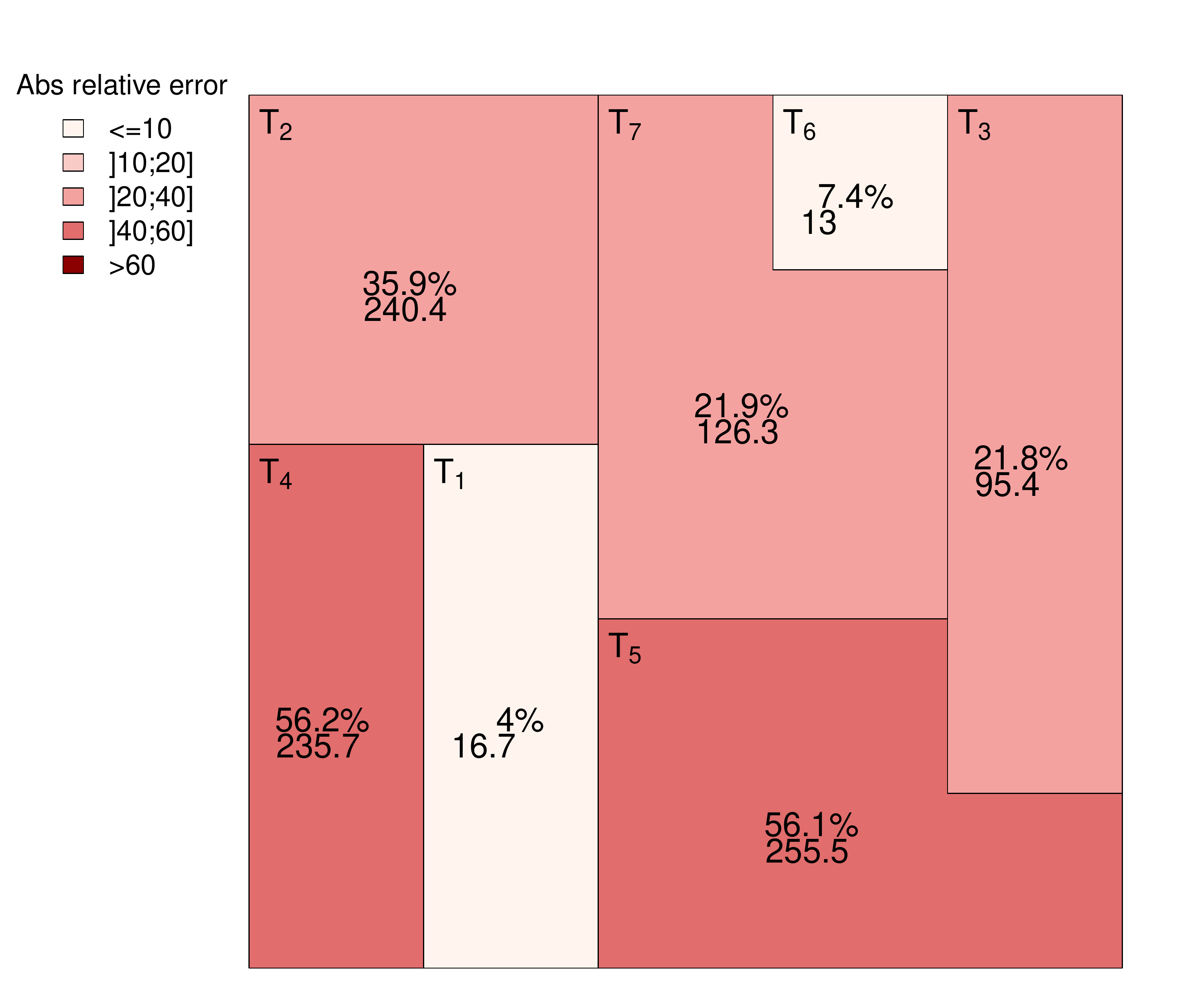}}
\subfloat[DAW for $Y_2$]{\includegraphics[width = 1.5in]{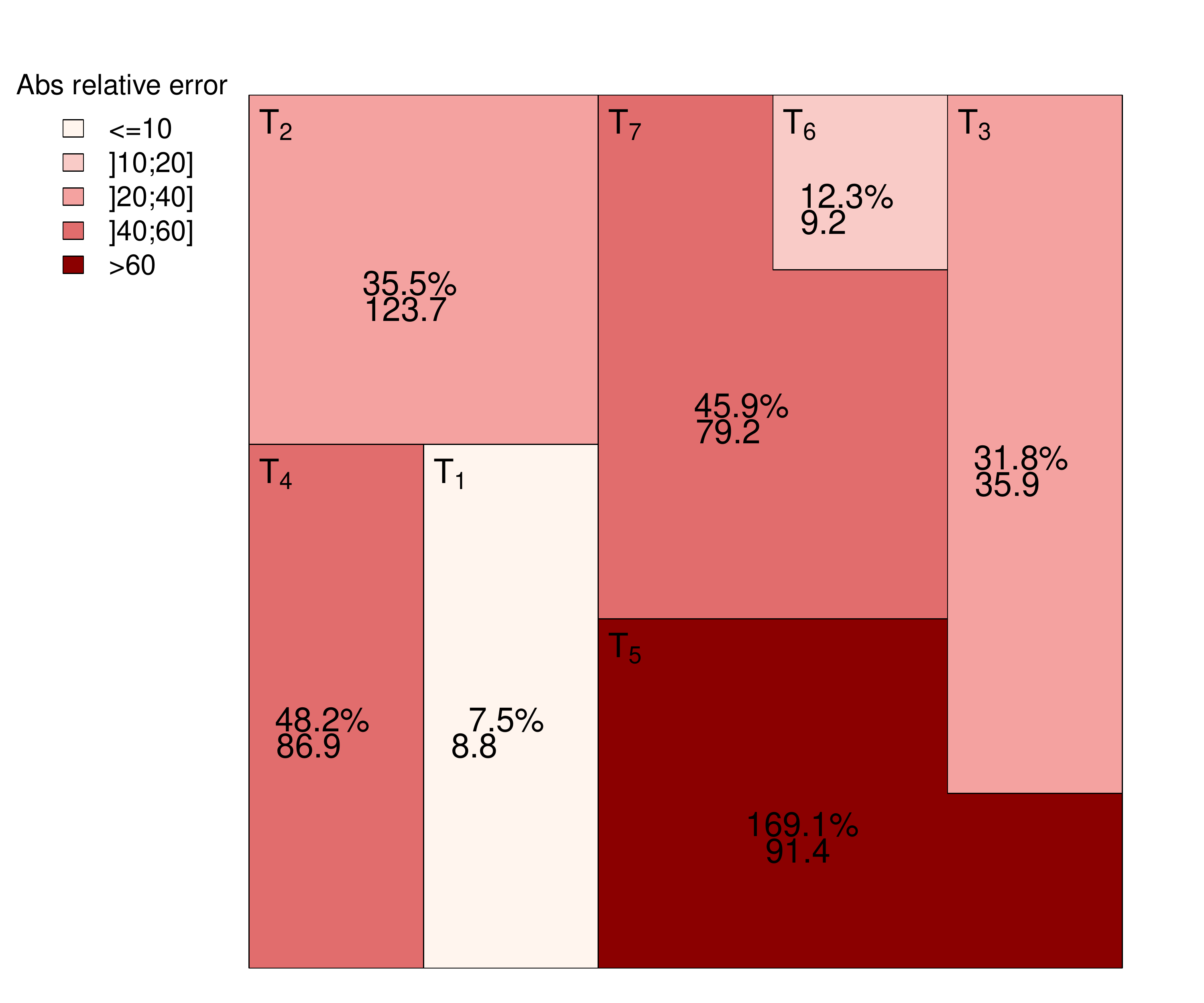}}
\subfloat[DAX for $Y_2$]{\includegraphics[width = 1.5in]{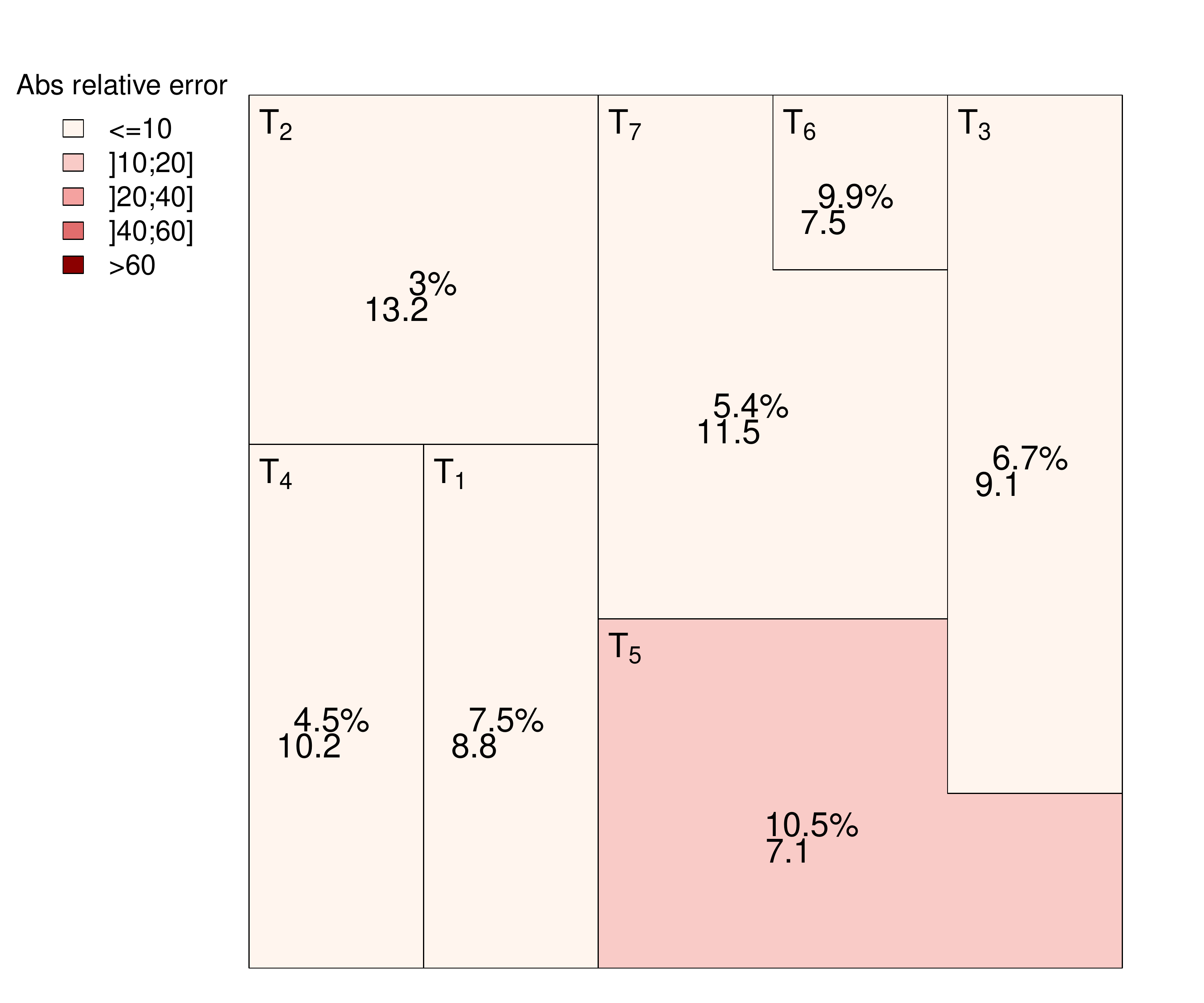}}
\caption{Toy example. Comparison of areal weighting interpolation and dasymetric method.}
\label{resulsDAWDAX}
\end{center}
\end{figure}

In order to illustrate our findings, we use a simulated toy example. We  intentionally drop the assumption that targets are nested in sources which was made for mathematical convenience and this will allow us to test the robustness of the results with respect to that assumption. On a square grid with 25 cells, we design three sources and seven targets as unions of cells. On Figure \ref{toyex1}, we see the design of sources and targets together with the cell counts for two target variables $Y_1$ and $Y_2$ and one auxiliary variable $X$ (for one particular draw). To generate $X$, we simulate a Poisson point process with an inhomogeneous intensity. We then recover the counts at the cell level to get the auxiliary information. The two target variables are then generated according to their relationship with the auxiliary variable (model (\ref{intens})) and the source values are  obtained by aggregation of cells. The true value of two target variables  is also shown at target level for accuracy comparison for one particular draw. For $Y_1$, we use the set of parameters $\alpha=80$ and  $\beta=1$ and for $Y_2$ we use $\alpha=0$ and $ \beta=1$ so that the area has a strong impact on $Y_1$ and that $Y_2$ is only driven by $X$. Conditionally upon one draw of $X$ (for which we
observe 1011 points), we draw 1000 repetitions of $Y_1$ and $Y_2$ and present the relative error and the error at target level on Figure \ref{resulsDAWDAX}.

The accuracy criterion is  an average of the error over all 1000 draws. We  present the relative error and the error at target level on Figure \ref{resulsDAWDAX}.

Figure \ref{resulsDAWDAX} is in agreement with the theoretical results: areal weighting interpolation is better than dasymetric for $Y_1$ for which the areal effect dominates and worse for $Y_2$ for which the auxiliary information effect driven by $X$ dominates. We can see that $Y_2$ is less homogeneous than  $Y_1$ on Figure \ref{toyex1} (a) and  (c): at the right-bottom target zone $T_5$ has very small  $Y_2$ counts. The dasymetric predictor is therefore very good for $Y_2$  (Figure \ref{resulsDAWDAX} (d)). To be more precise, we compute the ratio of the two average errors at target level for the two methods and it shows that areal weighting is best  for $Y_1$ with a ratio of square root of errors of ${201 \over 452} = 44\%$, whereas dasymetric is  best for $Y_2$  with  ratio of square root of errors of ${26 \over 197} = 13\%$. Table \ref{MSEtoyex1} reports the square root of the overall regional error (from formula (\ref{overall})) for the three methods: areal weighting, dasymetric and Poisson regression. For $Y_1$, the regression method is best, for $Y_2$ dasymetric is best because the impact of $X$ is strong (almost no areal effect).

 \begin{table}
 \begin{center}
\caption {\bf Square root of overall regional errors}\label{MSEtoyex1}
\begin{tabular}{|m {6cm}|c|c|}
\hline
\centering  Methods&$Y_1$&$Y_2$\\
\hline
Areal weighting interpolation&201&197\\
\hline
 Dasymetric&452&26\\
\hline
 Regression&55&33\\
\hline
\end{tabular}
\end{center}
\end{table}

For the practitioner, an important question is to be able to guess which method will perform better in a given situation. We might believe that a good correlation between $Y$ and $X$ is a sign that dasymetric based on $X$ will perform better than areal weighting. However in our case the correlation between $Y_1$ and $X$ is $0.94$ and the correlation between $Y_2$ and $X$ is $0.98$ which shows that this is a bad idea to rely on correlation. We could look at a measure of homogeneity to predict that areal weighting is the best method: in our case, the Gini coefficient of $Y_1$ is $0.14$ and of $Y_2$ is $0.40$ which goes in that direction. As we have seen in Theorem \ref{TheoDiffErDaxDaw}, the sign of the imbalance index at source level $\Delta_S$ determines which method is best (see Table \ref{BalanceToyex1}).  

To be more precise, let us examine the results of Table  \ref{ErrorInterLevel} in comparison with Theorem \ref{TheoDiffErDaxDaw}. Theorem \ref{TheoDiffErDaxDaw} shows that the difference of errors at target level is influenced by three factors: the mean number of points of the source, the imbalance of the source and the inhomogeneity of the auxiliary information of the given target. Let us look at each impact. For the influence of the inhomogeneity of the auxiliary variable, we compare targets of a  source  $S_1$ for example.  The first two impacts are constant (1164 points and 0.10 imbalance), and we see that the more homogenous the auxiliary information is (in increasing order $A_{16}, A_{13}, A_{17}, A_{12}$), the more distant the two methods are (18, 20, 1728, 2529 respectively). To examine the impact of the imbalance, we consider intersection zones $A_{17}$ and $A_{33}$ which are nested within two sources with  similar number of points ($S_1$ with 1164 points and $S_3$ with 1168 points respectively). Even though their inhomogeneity are not very different  (0.12 vs 0.10), the difference of the errors on $A_{33}$  is 3.7 times larger than on $A_{17}$. This fact is explained by the distance between the two imbalances of $S_1$ and $S_3$ (0.10 vs 0.51). The last but not least effect is the number of points of the source zones. The comparison between $A_{24}$ and $A_{35}$ shows its impact: they have similar inhomogeneities (0.20 vs 0.19), not too different imbalances (0.41 vs 0.51), but the difference in errors for $A_{35}$ (25707) is three times larger than the difference in errors for  $A_{24}$ (7938) and this is linked to the discrepancy in the mean numbers of points (1168 vs 679). As Table \ref{ErrorInterLevel} shows, the combination of the three impacts is very complex. In other words, choosing between the areal weighting interpolation and the dasymetric is not an easy problem.

\begin{table}
 \begin{center}
 \caption{\bf Imbalance index at source level}\label{BalanceToyex1}
\begin{tabular}{|c|c|c|c|}
\hline
Source zones&$S_1$&$S_2$&$S_3$\\
\hline
$Y_1$& 0.10& 0.41& 0.51\\
\hline
$Y_2$& -1& -1& -1\\
\hline
\end{tabular}
\end{center}
\end{table}

{\small
\begin{table}[h]
\hspace{-2.5cm}
\begin{tabular}{|c|c|c|c|c|c|c|c|c|c|c|c|c|c|c|}
\hline
\multicolumn{1}{|c|}{Sources}&\multicolumn{4}{c|}{$S_1$}&\multicolumn{3}{c|}{$S_2$}&\multicolumn{6}{c|}{$S_3$}\\
\hline
$\E(Y_1)$&\multicolumn{4}{c|}{1164}&\multicolumn{3}{c|}{679}&\multicolumn{6}{c|}{1168}\\
\hline
Imbalance&\multicolumn{4}{c|}{0.10}&\multicolumn{3}{c|}{0.41}&\multicolumn{6}{c|}{0.51}\\
\hline
\multicolumn{1}{|c|}{Intersections}&$A_{12}$&$A_{13}$&$A_{16}$&$A_{17}$&$A_{21}$&$A_{24}$&$A_{25}$&$A_{31}$&$A_{32}$&$A_{33}$&$A_{34}$&$A_{35}$&$A_{37}$\\
\hline
\multicolumn{1}{|c|}{$|{|A|\over |S|}-{X_A \over X_S}|$}&0.14&0.01&0.01&0.12&0.03&0.20&0.18&0.02&0.17&0.10&0.06&0.19&0.09\\
\hline
$ \Er^{DAX}-\Er^{DAW}$& 2529&20&18&1728&135&7938&5999&199&19858&6435&16817&25707&2240\\
\hline
$\Re^{DAX}/\Re^{DAW}$& 1.5&1.3&1.1&1.4&1.8&5.4&5.4&3.1&8.8&7.5&8.7&8.8&6.0\\
\hline
\end{tabular}
\caption{\bf Errors at intersection level for $Y_1$}\label{ErrorInterLevel}
\end{table}
}

\section{Relative accuracy of the other methods: asymptotic assessment}
\label{acc2}
Let us now try to extend the comparison to  the Poisson regression method. This cannot be done anymore by finite distance methods and so we introduce an asymptotic framework.
Model \eqref{model} yields at source level
\begin{equation}
Y_s\sim \P(\alpha |S_s|+\beta x_s)
\end{equation}
where $x_s=\sum_{t: t\cap s \not= \emptyset} x_{st}$. Besides the Poisson regression predictor defined by \eqref{RegPredictor}, inspired by Theorem \ref{composite}, we propose a new predictor called scaled Poisson regression predictor defined as follows
\begin{equation}\label{SclPredictor}
\hat{Y}^{ScR}_{st}=\dfrac{\hat{\alpha}|A_{st}| + \hat{\beta}x_{st}}{\hat{\alpha}|S_s| + \hat{\beta}x_{s}}Y_S,
\end{equation}
\noindent where $\hat\alpha$ and $\hat\beta$ are the estimators of $\alpha$ and $\beta$ obtained through the Poisson regression at source level. Note that if model \eqref{intens} contains only one of the two effects (that of $X$ for example), then it is easy to see that
the predictor of the scaled regression method coincides with the dasymetric method (corresponding to $X$):
\begin{align*}
\hat{Y}^{ScR}_T={\hat{\beta}x_T\over \hat{\beta}x_S}Y_S=\hat{Y}^{DAX}_T.
\end{align*}
In section \ref{est}, we establish the asymptotic properties of the estimators $\hat\alpha$ and $\hat \beta$ and these results will enable us to compare the predictors in section \ref{predict}. Section \ref{subsectionToyexample} illustrates these results on a toy example. 

\subsection{Estimators of the regression coefficients}
\label{est}
 In this section, we adapt proofs from Fahrmeir and Kaufmann (1985)  to establish the consistency and asymptotic normality of the estimators $\hat{\alpha}, \hat{\beta}$. We first need to describe an asymptotic framework. To be realistic, we assume that the whole region $\Omega$ is fixed and that the number of source zones $n_S$ (hereafter denoted by $n$) increases to infinity. In this section, the source zones will be denoted by $S_{n,i}: i=1,2,...,n$  and $\Omega = \cup_i S_{n,i}$.
Because of the extensive property of $X$, we also assume a similar property of $X_{n,i}$: the total auxiliary information on the region $\Omega$ remains constant $x_\Omega = \sum_i x_{n,i}.$ In order to get a consistent regression however we need the amount of information at source level to increase and we thus assume that the intensity of $Y$ increases with a rate $k_n \longrightarrow \infty$ so that

$$Y_{A}\sim \P(\alpha \widetilde{|A|}+\beta \tilde{x}_A)$$
where $\widetilde{|A|}=k_n|A|, \tilde{x}_A=k_n x_A$.

 \noindent Let $\gamma=\colvec{\alpha}{\beta}, Z_A=\colvec{|A|}{x_A}, Z_{n,i}=\colvec{|S_{n,i}|}{x_{n,i}}$. With these notations we have $\lambda_A=\gamma' Z_A, \tilde{Z}_A= \colvec{\widetilde{|A|}}{\tilde{x}_A}=k_n Z_A$   and  $Y_A \sim \P(k_n \lambda_A)$. The true value of the parameter $\gamma$ will be denoted by $\gamma_o=\colvec{\alpha_o}{\beta_o}$.

\noindent The log likelihood function $l_n(\gamma)$, the score function $s_n (\gamma)$ and the information matrix $F_n (\gamma)$ are then given by
\begin{align*}
l_n(\gamma) = \sum_{i=1}^n y_{n,i} \ln (\gamma' \tilde{Z}_{n,i})-\gamma' \tilde{Z}_{n,i}-\ln(y_{n,i}!)
\end{align*}

\begin{align*}
s_n(\gamma)&=\dfrac{\partial l_n(\gamma)}{\partial \gamma}  = \sum_{i=1}^n \dfrac{\tilde{Z}_{n,i}}{\gamma' \tilde{Z}_{n,i}} y_{n,i} - \tilde{Z}_{n,i}\\
F_n(\gamma)&= cov_\gamma (s_n(\gamma))=\sum_{i=1}^n \dfrac{\tilde{Z}_{n,i}\tilde{Z}'_{n,i}}{\gamma' \tilde{Z}_{n,i}}
\end{align*}
Differentiation of the score yields
\begin{align*}
H_n(\gamma)&=-\dfrac{\partial s_n(\gamma)}{\partial \gamma}=\sum_{i=1}^n \dfrac{\tilde{Z}_{n,i}\tilde{Z}'_{n,i}}{(\gamma' \tilde{Z}_{n,i})^2}y_{n,i}
\end{align*}
It is easy to see that $\E_\gamma(s_n(\gamma))=0, \E_\gamma(H_n(\gamma))=F_n(\gamma)$. We further simplify the notations and use $s_n, F_n, H_n, \E$  instead of $s_n(\gamma_o), F_n(\gamma_o), H_n(\gamma_o), \E_{\gamma_o}$.  It is clear that the matrix $H_n$ is positive definite and therefore the log likelihood function is concave which leads to a unique minimum. In the sequel, we also need the square root $F_n^{1/2}$ of the symmetric matrix $F_n$, i.e. $F_n^{1/2}F_n^{1/2}=F_n$.

\noindent Our asymptotic framework differs from that of Fahrmeir and Kaufmann (1985) in the sense that at each step they have one new observation whereas in our case at each step all observations are new and we have one more than at the previous step. For this reason,  we modify slightly their conditions and assume that
\begin{itemize}
\item[(C1)] $\{\tilde {Z}_{n,i}\} \subset \Z \,\forall n,i$ with $\Z$ is a compact set.
\item[(C2)] $\lambda_{min} (\sum_i \tilde {Z}_{n,i}' \tilde {Z}_{n,i}) \to \infty$ as $n \to \infty$ where $\lambda_{min} (W)$ denotes the minimum eigenvalue of the matrix $W$.
\end{itemize}

\noindent Condition (C1) is satisfied if  there exists two positive numbers $c_1, c_2$ (note that $||\tilde{Z}_{n,i}|| \not= 0$) s.t.
\begin{equation}\label{conditionC1}
c_1 <||\tilde{Z}_{n,i}||< c_2
\end{equation}
\noindent In that case, the number of source zones increases with the rate of growth of the intensity at a similar rate and the number of points in one source zone is quite stable during the change process.

\noindent Under these conditions, we get the following asymptotic behavior for the Poisson regression coefficients.

\begin{theorem}\label{ConsitencyNormality}
Under conditions (C1) and (C2), the following statements holds for  the Poisson regression estimator $\hat{\gamma}_n$ of $\gamma$
\begin{itemize}
\item[(i)]$\hat{\gamma}_n \to_p \gamma_o$  (weak consistency)\\
\item[(ii)] $F_n^{1/2} (\hat{\gamma}_n - \gamma_o) \to_d {\cal N}(0,\I)$ (asymptotic normality)
\end{itemize}
\end{theorem}


\noindent In the next section, we use these results to study the asymptotic behavior of the predictors.
\subsection{Predictors}
\label{predict}
In this section, we consider the asymptotic properties of the following two predictors: the regression predictor \eqref{RegPredictor} and the scaled regression predictor \eqref{SclPredictor}. We prove that the scaled regression predictor is asymptotically as accurate as the unfeasible composite predictor. We also compare these two methods with areal weighting interpolation and dasymetric interpolation.

The first proposition is concerned with the pycnophylactic property, which is of interest in the areal interpolation literature. It shows that, whereas it is satisfied  by the scaled Poisson regression predictor, it is not satisfied at source level by the ordinary Poisson regression predictor but only at region level.

\begin{proposition}\label{ProPycno}
The scaled Poisson regression predictor satisfies the pycnophylactic property at source level.  The ordinary Poisson regression predictor is  pycnophylactic at region level  and asymptotically pycnophylactic at source level.
\end{proposition}
\noindent To prove proposition \ref{ProPycno}, we need the following asymptotic normality result for the target variables

\begin{equation}\label{app}
\dfrac{Y_{ni}-\gamma_o' \tilde{Z}_{n,i}}{\sqrt{\gamma_o' \tilde{Z}_{n,i}}}\to_d \N(0,1)
\end{equation}
\noindent We now turn attention to the asymptotic behavior of the prediction error for the ordinary Poisson regression predictor.

\begin{theorem}\label{TheNormalityREG}
The asymptotic normality of the prediction error of the Poisson regression predictor at source level is given by
$$\dfrac{\hat{Y}_{ni}^{REG}-Y_{ni}}{\sqrt{\gamma_o'\tilde{Z}_{n,i}}} \to_d \N(0,1)$$

\noindent If we also assume  a lower bound for $\tilde{Z}_{T}$, the following similar result at the target level holds
$$\dfrac{\hat{Y}_{T}^{REG}-Y_{T}}{\sqrt{\gamma_o'\tilde{Z}_{T}}} \to_d \N(0,1)$$
\end{theorem}

\noindent The next result is about the quadratic prediction error and relative prediction error of the Poisson regression predictor.

\begin{theorem}\label{TheErrReg}
For any $\eta >0$, there exists a sequence of sets $\{Q_i\}_i: \Pr(Q_i) \to 1$ such that
\begin{align*}
-\eta +{\gamma}_o'\tilde{Z}_T <\E(\hat{Y}^{REG}_T-Y_T)^2\1_{Q_i}<\eta +{\gamma}_o'\tilde{Z}_T
\end{align*}

\noindent If the number of target zones contained in one source zone $S_{n,i}$ is bounded,  the relative error at source level can be approximated by
\begin{equation}\label{approRelativeErrorREG}
\Re^{REG}_{n,i}\approx {1 \over \sqrt{\E(Y_{n,i})}}={1 \over \sqrt{\alpha_o \widetilde {|S_{n,i}|} + \beta_o \tilde{x}_{n,i}}}
\end{equation}
\end{theorem}

\noindent Because $Var(Y_T)=\E(Y_T)={\gamma}_o'\tilde{Z}_T$, this theorem says that the quadratic prediction error of the regression predictor is asymptotically equivalent to the variance of the underlying process. Equation \eqref{approRelativeErrorREG} shows that the relative error of the regression predictor is going to be small when the number of points on a source zone is large. However, this number being bounded by condition (C1), this relative error cannot converge to zero in this framework.

%

\noindent Let us now turn attention to the difference between the relative prediction errors of the Poisson regression method and that of the areal weighting and the dasymetric methods.
If the target zones are nested within the source zones and the number of target zones contained in one source is bounded, we get the following approximation at source level for the differences between the relative errors of the methods when $\E(Y_{n,i})$ are large:

\begin{align}\label{diffapproRelEr}
[(\Re^{REG}_{n,i})^2-(\Re^{DAW}_{n,i})^2]
&\approx -(1 + \Delta_{n,i})^2 \sum_T (\dfrac{{|T|}}{{|S_{n,i}|}}-\dfrac{{x}_T}{{x}_{n,i}})^2\\
[(\Re^{REG}_{n,i})^2-(\Re^{DAX}_{n,i})^2]
&\approx -(1 - \Delta_{n,i})^2 \sum_T (\dfrac{{|T|}}{{|S_{n,i}|}}-\dfrac{{x}_T}{{x}_{n,i}})^2\\
\end{align}

\noindent This result shows that, among the three methods: areal weighting, dasymetric and Poisson regression,  regression outperforms the other two methods asymptotically (negative sign). However, from the proof in the annex, we can see that if $(\dfrac{\widetilde{|T|}}{\widetilde{|S_{n,i}|}}-\dfrac{\tilde{x}_T}{\tilde{x}_{n,i}})=0$ then the regression is less accurate than areal weighting and dasymetric asymptotically so that none of them is always dominant.

\noindent  For  areal weighting and dasymetric predictors, we have seen that if one method is better on one target, then it is also true on all targets contained on the same source zone. The difference between the accuracy of the regression method and the other two methods depends on the difference of ratios ${|T| \over x_T}-{|S| \over x_S}$: the higher this difference, the larger the difference between regression and the other two.

\noindent The fact that the regression predictor doesn't satisfy the pycnophylactic property is not surprise but the fact that it does satisfy this property on the whole region is interesting. The idea of scaling to obtain the pycnophylactic property can be found also in  Yuan et. al. (1997) for ordinary linear regression without theoretical justifications; we have extended it to the Poisson regression case and provided some theoretical motivation for it.

\noindent We now turn attention to the scaled regression and prove it is better than the unscaled one and that
 its accuracy can be approximated by that of the unfeasible composite predictor.

\noindent The first lemma proves an asymptotic equivalence between the scaled regression predictor and the unfeasible composite predictor.

\begin{lemma}\label{LeScRCom}
For any target $T$,
\begin{equation}
\hat{Y}^{ScR}_T-\hat{Y}^{C}_T \to_p 0
\end{equation}
\end{lemma}

\noindent The next result is about the quadratic prediction error  of the scaled Poisson regression predictor.
\begin{theorem}\label{TheScR}
For any $\eta >0$, there exists a sequence of sets $\{Q_i\}_i: \Pr(Q_i) \to 1$ such that
$$-\eta +\tilde{Z_T}\gamma_o - {(\tilde{Z}_T \gamma_o)^2 \over \tilde{Z}_{n,i}\gamma_o}<\E(\hat{Y}^{ScR}_T-Y_T)^2\1_{Q_i}< \eta +\tilde{Z_T}\gamma_o - {(\tilde{Z}_T \gamma_o)^2 \over \tilde{Z}_{n,i}\gamma_o}
  $$
\end{theorem}
Since $\Er^{C}_T=\tilde{Z_T}\gamma_o - {(\tilde{Z}_T \gamma_o)^2 \over \tilde{Z}_{n,i}\gamma_o}$, this theorem shows that the quadratic prediction error of the scaled regression predictor is asymptotically equivalent to the one of the composite. Consequently, the scaled regression method is the best among the areal weighting, the dasymetric and the regression predictors.

\subsection{Accuracy: simulation assessment with a toy example}\label{subsectionToyexample}

\begin{figure}[h!]\label{cellsandsources}
\begin{center}
\subfloat[Cells]{\includegraphics[width = 2.5in]{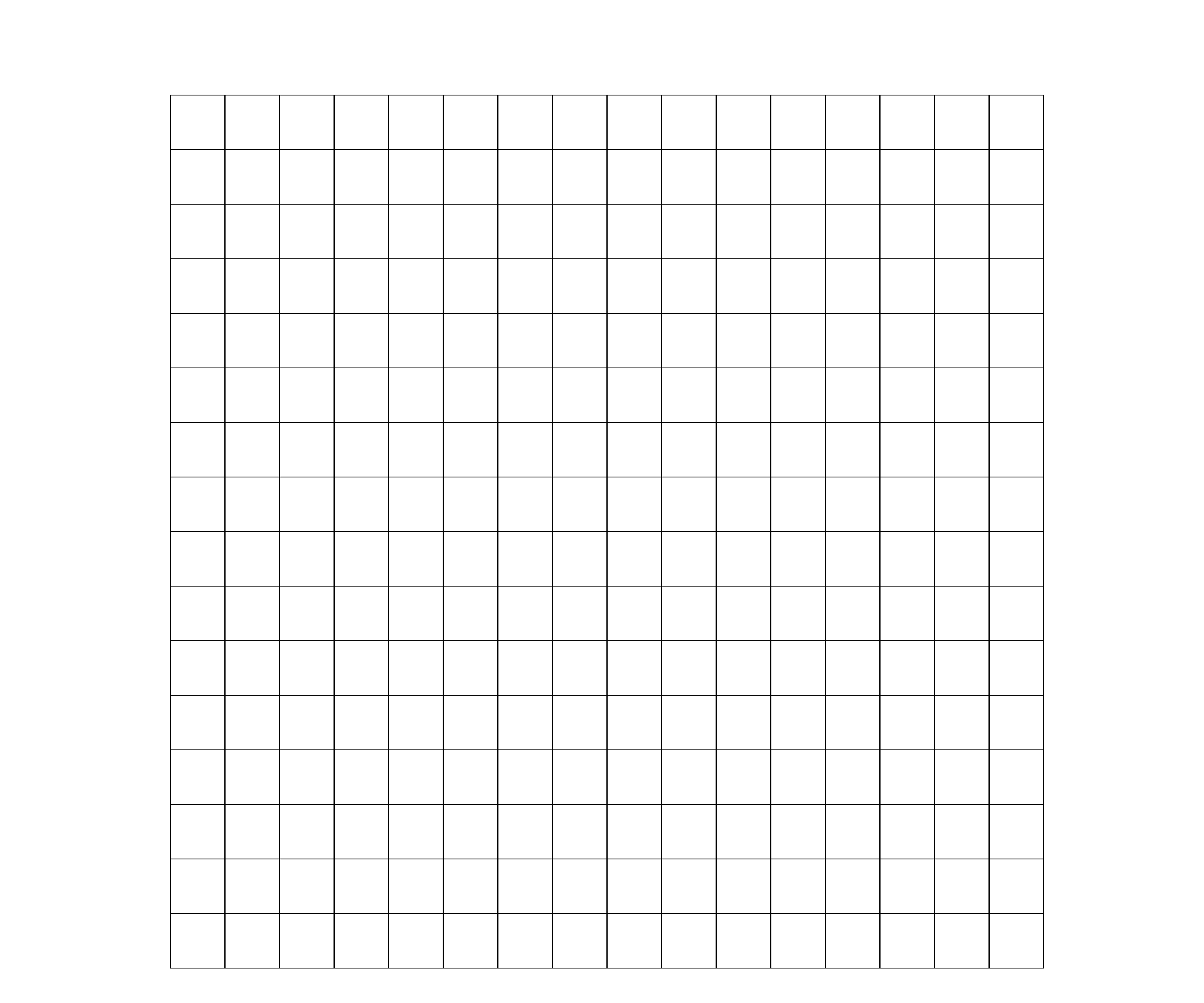}}
\subfloat[Sources 1]{\includegraphics[width = 2.5in]{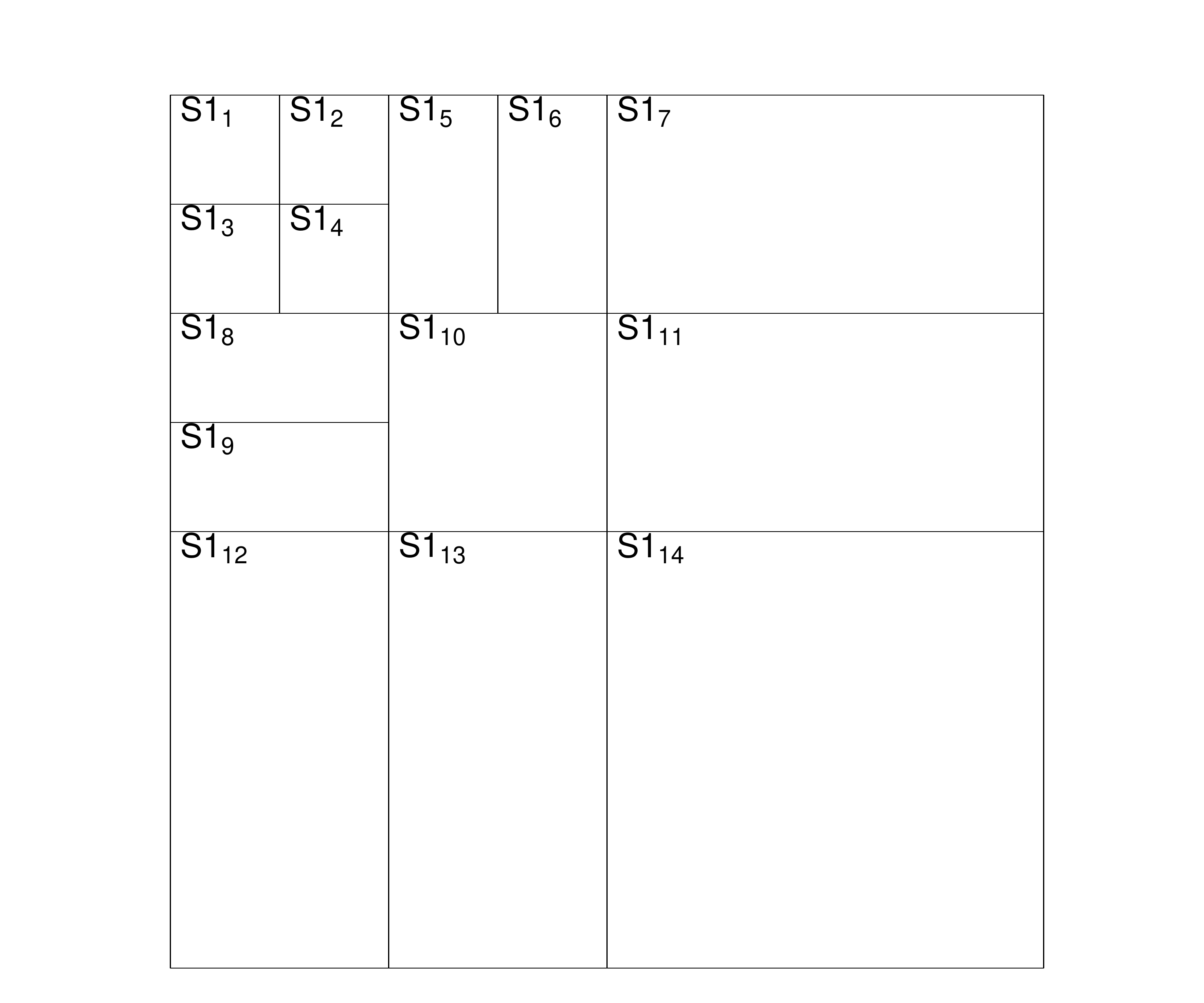}}\\
\subfloat[Sources 2]{\includegraphics[width = 2.5in]{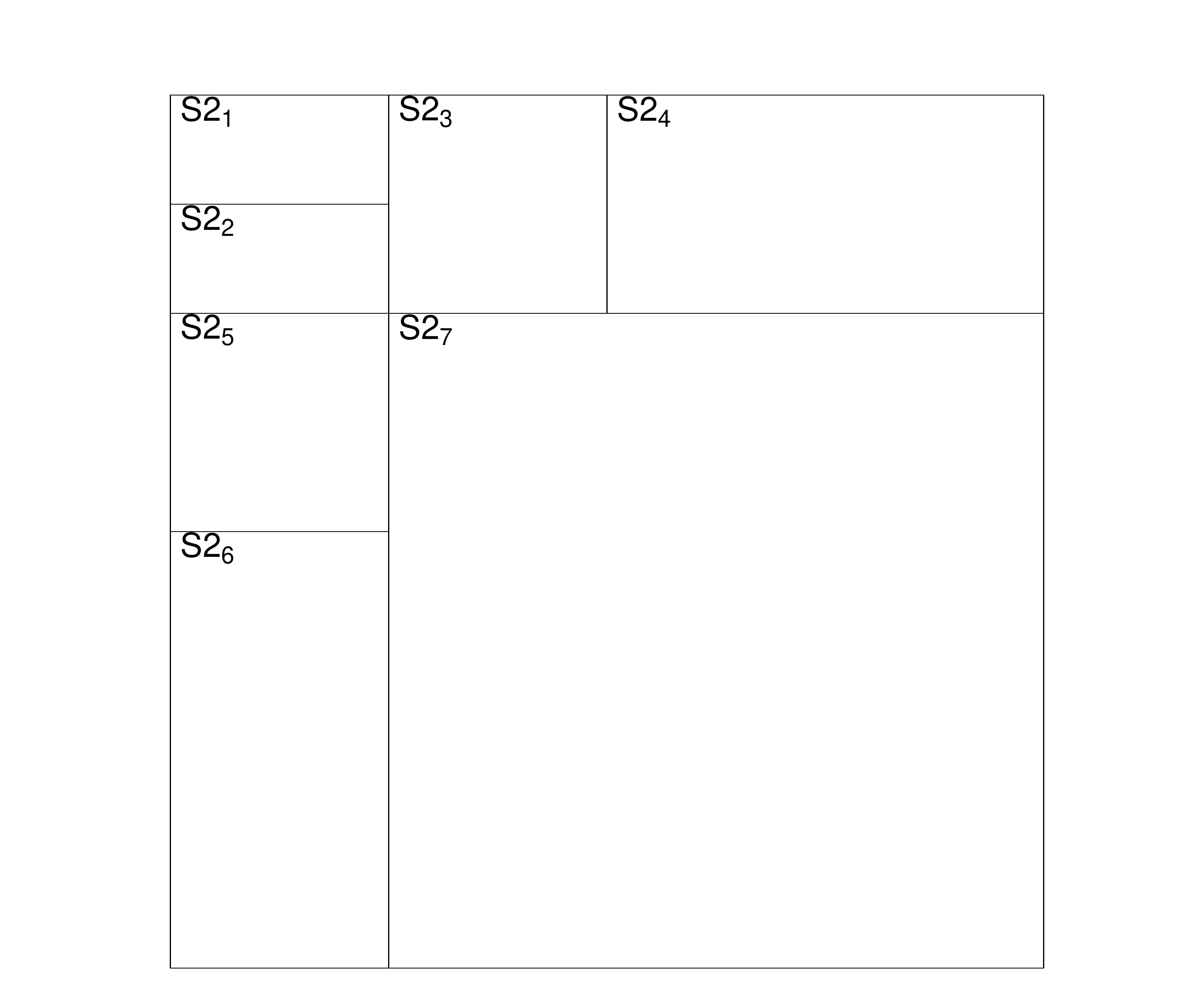}}
\subfloat[Sources 3]{\includegraphics[width = 2.5in]{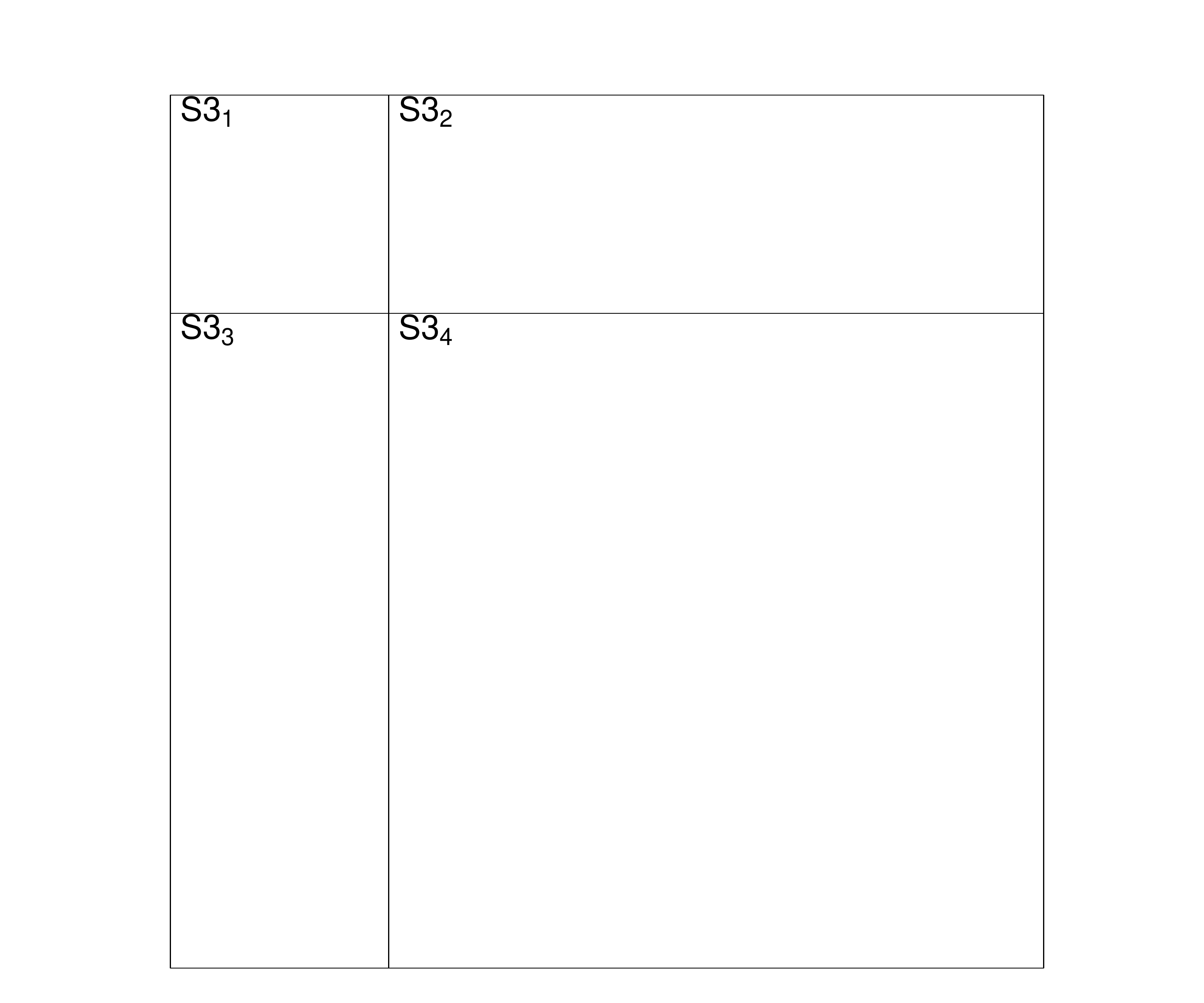}}
\caption{Spatial polygons}
\end{center}
\end{figure}

 We devise a simple simulation to illustrate these results. On a square region $\Omega$ with $16 \times 16$ cells,  we build three systems of sources with respectively 14, 7 and 4 sources (see Figure \ref{cellsandsources}). We simulate two Poisson point patterns (our auxiliary information) with an expected overall number of points of $100,000$:  $X_1$  is very inhomogeneous (Gini coefficient of cell counts of  0.74 with 100,247 points)  and $X_2$  is very homogeneous (Gini coefficient of cell counts of  0.03 with  100,008 points).

\noindent Target variables are then generated following model (\ref{model}). For each of the auxiliary variables, we choose four couples of coefficients $\alpha, \beta$ to study the  effects of imbalance so that we get eight different target variables. Table \ref{imb} reports  the minimum, maximum and average imbalance for each case and for each system of source zones.

 {\small
\begin{table}
\caption{\bf Imbalances}\label{imb}
\vskip 0.5cm
\hspace{-1cm}
\begin{tabular}{|c|c|r|r|r|}
\hline
\bf Case&\bf Sources&\bf Min& \bf Mean & \bf Max\\
\hline
 $\alpha=100$&14 sources& -0.92& -0.40& 0.96
  \\
\cline{2-5}
$ \beta=1$&7 sources&  -0.90&-0.59&0.64
   \\
\cline{2-5}
$ X_1$&4 sources& -0.87&-0.44& 0.64
 \\
 \hline\hline
 $\alpha=600$&14 sources&  -0.62& 0.11&0.99
  \\
\cline{2-5}
$ \beta=1$&7 sources&  -0.53&-0.05&0.93
  \\
\cline{2-5}
$ X_1$&4 sources& -0.43& 0.07& 0.93
 \\
\hline \hline
 $\alpha=1000$&14 sources& -0.43&0.29& 1
  \\
\cline{2-5}
$ \beta=1$&7 sources&   -0.33&0.17&0.96
  \\
\cline{2-5}
$ X_1$&4 sources&   -0.21& 0.26& 0.96
\\
\hline \hline
 $\alpha=1000$&14 sources& 0.6& 0.86& 1
  \\
\cline{2-5}
$ \beta=0.1$&7 sources&  0.67&0.84& 1
  \\
\cline{2-5}
$ X_1$&4 sources&   0.74& 0.86& 1
\\
\hline
\end{tabular}
\quad
\begin{tabular}{|c|c|r|r|r|}
\hline
\bf Case&\bf Sources&\bf Min& \bf Mean & \bf Max\\
\hline
 $\alpha=100$&14 sources&  -0.63& -0.6& -0.58
  \\
\cline{2-5}
$ \beta=1$&7 sources&  -0.61& -0.6&-0.59
   \\
\cline{2-5}
$ X_2$&4 sources&  -0.6&-0.6&-0.59
 \\
\hline \hline
 $\alpha=600$&14 sources& 0.16& 0.21& 0.22
  \\
\cline{2-5}
$ \beta=1$&7 sources&  0.18& 0.2& 0.22
  \\
\cline{2-5}
$ X_2$&4 sources&0.19& 0.2& 0.22
\\
\hline \hline
 $\alpha=1000$&14 sources& 0.39& 0.43& 0.45
 \\
\cline{2-5}
$ \beta=1$&7 sources& 0.41& 0.43& 0.44
  \\
\cline{2-5}
$ X_2$&4 sources& 0.42& 0.43& 0.44
\\
\hline \hline
 $\alpha=1000$&14 sources&0.92& 0.92& 0.93
  \\
\cline{2-5}
$ \beta=0.1$&7 sources&0.92& 0.92& 0.93  \\
\cline{2-5}
$ X_2$&4 sources&0.92& 0.92& 0.93 \\
\hline
\end{tabular}
\end{table}
}

  We then apply the four considered methods (areal weighting, dasymetric, Poisson regression and scaled Poisson regression) to transfer the data from each of the three systems of source zones to cell level.  For each case, we generate the data 1000 times, and calculate prediction errors for each method and each iteration. Table \ref{errorToyex2} (respectively Table \ref{relativeErrorToyex2}) reports the average absolute square root of prediction errors (respectively the average absolute square root of relative prediction errors). The two tables also present the mean of the target variables at region level $\E(Y_\Omega)=\alpha |\Omega|+ \beta x_\Omega$  (because it appears in Theorem \ref{TheErrReg}) and  the theoretical composite prediction error as a benchmark (see Theorem \ref{TheScR}).

\begin{center}
\begin{table}
\caption{\bf Square root of prediction errors.}\label{errorToyex2}
\vskip 0.5cm
\begin{tabular}{|>{\centering}p{2cm}|r|r|c|r|r|r|r|}
\hline
\multicolumn{1}{|c|}{\bf Methods}  & {\centering{\bf $\sqrt{\E(Y_\Omega)}$}}& {\centering \textbf{Sources}} &{\centering \textbf{DAW }} & {\centering \textbf{DAX}} &{\centering \textbf{REG}}&{\centering \textbf{ScR}}&{\centering{\bf Composite}}\\
\hline
 $\alpha=100$&\multirow{3}{*}{ 354.7
}&14 sources&6580.5 &1621.6  &353.0  &333.1 & 332.1  \\
\cline{3-8}
$ \beta=1$&&7 sources&6962.7 &2087.7  &352.1 & 341.3 &341.2
 \\
\cline{3-8}
$ X_1$&&4 sources&7215.7& 2090.6 & 352.1  &347.6 &347.8\\
\hline
\hline
 $\alpha=600$&\multirow{3}{*}{ 503.8
}&14 sources&6589.6 &9532.5&  500.6&  481.9& 482.1
\\
\cline{3-8}
$ \beta=1$&&7 sources& 6971.3& 12363.7 &  502.9 &  493.4 & 491.7 \\
\cline{3-8}
$ X_1$&&4 sources&  7225.3 &12374.7  & 502.5  & 498.8& 497.4  \\
\hline
\hline
 $\alpha=1000$&\multirow{3}{*}{ 596.9
}&14 sources& 6597.3& 15878.2&   594.0 &  574.5& 574.2\\
\cline{3-8}
$ \beta=1$&&7 sources&  6982.2& 20595.2&   595.6&   586.5 &  584.6
\\
\cline{3-8}
$ X_1$&&4 sources&7229.6 &20614.5 &  594.4  & 590.8 & 590.3
\\
\hline
\hline
 $\alpha=1000$&\multirow{3}{*}{ 515.8
}&14 sources&826.7 &15875.6 &  513.6  & 500.7 &500.9 \\
\cline{3-8}
$ \beta=0.1$&&7 sources&   861.9& 20592.8&  515.0&  509.4 &  508.3
  \\
\cline{3-8}
$ X_1$&&4 sources&   883.5 &20612.4  & 514.5  & 512.3  &   511.6   \\
\hline
\hline
 $\alpha=100$&\multirow{3}{*}{ 354.4
}&14 sources&458.8& 353.5& 356.6& 348.3&344.5 \\
\cline{3-8}
$ \beta=1$&&7 sources& 469.0& 358.6& 357.7& 354.2 &349.5
 \\
\cline{3-8}
$ X_2$&&4 sources&474.2 &361.3 &359.6 &358.2& 351.6
 \\
\hline
\hline
$\alpha=600$&\multirow{3}{*}{ 503.6
}&14 sources&573.9 &677.4& 504.9& 492.9 &489.6 \\
\cline{3-8}
$ \beta=1$&&7 sources& 587.3& 690.0& 508.1&503.1& 496.6  \\
\cline{3-8}
$ X_2$&&4 sources&591.2& 697.7 &509.1& 507.2& 499.6\\
\hline
\hline
 $\alpha=1000$&\multirow{3}{*}{ 596.7
}&14 sources&654.8& 969.9& 599.0 &585.0 & 580.1 \\
\cline{3-8}
$ \beta=1$&&7 sources& 665.5& 992.6& 601.7& 595.7 &  588.4 \\
\cline{3-8}
$ X_2$&&4 sources&  671.8 &1002.6  &602.2 & 600.0 & 592.0 \\
\hline
\hline
$\alpha=1000$&\multirow{3}{*}{ 515.8
}&14 sources&502.9& 924.7 &518.6& 506.8 &501.4  \\
\cline{3-8}
$ \beta=0.1$&&7 sources&   510.3 & 947.4 & 520.4 & 515.5&  508.6  \\
\cline{3-8}
$ X_2$&&4 sources&  512.9 & 960.9&  522.2  &520.2&     511.7  \\
\hline
\end{tabular}
\end{table}
\end{center}

\begin{center}
\begin{table}
\vskip 0.5cm
\begin{tabular}{|>{\centering}p{2cm}|r|r|c|r|r|r|r|}
\hline
\multicolumn{1}{|c|}{\bf Methods}  & {\centering{\bf $\E(Y_\Omega)$}}& {\centering \textbf{Sources}} &{\centering \textbf{DAW }} & {\centering \textbf{DAX}} &{\centering \textbf{REG}}&{\centering \textbf{ScR}}&{\centering{\bf Composite}}\\
\hline
 $\alpha=100$&\multirow{3}{*}{ 125847
}&14 sources&  50.459   &    47.464  &      6.979   &     6.872& 6.873
 \\
\cline{3-8}
$ \beta=1$&&7 sources& 55.664   &    54.021 &       6.970  &      6.941&  6.951
 \\
\cline{3-8}
$ X_1$&&4 sources&    57.122    &   54.054 &       6.978 &       6.966&  6.969
\\
\hline
\hline
 $\alpha=600$&\multirow{3}{*}{ 253847
}&14 sources&  20.458    &   59.087      &  3.492  &      3.422 &   3.421
\\
\cline{3-8}
$ \beta=1$&&7 sources&   21.865  &     73.067 &       3.495   &     3.471&   3.466
 \\
\cline{3-8}
$ X_1$&&4 sources&     22.557  &     73.208&        3.490 &       3.480& 3.481
  \\
\hline
\hline
 $\alpha=1000$&\multirow{3}{*}{ 356247
}&14 sources&  14.859  &     62.210  &      2.813  &      2.754&  2.757
\\
\cline{3-8}
$ \beta=1$&&7 sources&   15.827      & 77.405 &       2.823&        2.801 & 2.794
\\
\cline{3-8}
$ X_1$&&4 sources&  16.324    &   77.589 &       2.820&        2.811&  2.808
\\
\hline
\hline
 $\alpha=1000$&\multirow{3}{*}{ 266025
}&14 sources&  4.297    &   72.322  &      3.093&        3.018 &3.020
\\
\cline{3-8}
$ \beta=0.1$&&7 sources&   4.433  &     89.662   &     3.100    &    3.069& 3.064  \\
\cline{3-8}
$ X_1$&&4 sources&   4.508   &    90.028 &       3.098  &      3.086&  3.082    \\
\hline
\hline
 $\alpha=100$&\multirow{3}{*}{ 125608
}&14 sources&  5.685   &     4.503    &    4.545&        4.438  & 4.391
\\
\cline{3-8}
$ \beta=1$&&7 sources& 5.799  &      4.569 &       4.559 &       4.514& 4.455
 \\
\cline{3-8}
$ X_2$&&4 sources&  5.859      &  4.603 &       4.581 &       4.564 & 4.482
 \\
\hline
\hline
$\alpha=600$&\multirow{3}{*}{ 253608
}&14 sources&  3.574   &     4.121    &    3.186   &     3.108 & 3.088 \\
\cline{3-8}
$ \beta=1$&&7 sources&  3.655   &     4.196  &      3.205   &     3.174  &3.134
  \\
\cline{3-8}
$ X_2$&&4 sources&  3.679   &     4.236     &   3.211  &      3.199& 3.153 \\
\hline
\hline
 $\alpha=1000$&\multirow{3}{*}{ 356008
}&14 sources&  2.918 &       4.084  &      2.692&        2.627& 2.606
  \\
\cline{3-8}
$ \beta=1$&&7 sources&   2.966      &  4.172  &      2.704 &       2.677 &2.645
 \\
\cline{3-8}
$ X_2$&&4 sources&   2.993     &   4.211     &   2.706 &       2.696&  2.661
  \\
\hline
\hline
$\alpha=1000$&\multirow{3}{*}{ 266001
}&14 sources&  3.022    &    5.133 &       3.118&        3.045&   3.015
 \\
\cline{3-8}
$ \beta=0.1$&&7 sources&    3.068   &     5.246     &   3.129 &       3.099&   3.059
  \\
\cline{3-8}
$ X_2$&&4 sources&    3.084   &     5.320   &     3.140  &      3.128 & 3.078
  \\
\hline
\end{tabular}
\caption{\bf Relative prediction errors (in percentages).}\label{relativeErrorToyex2}
\end{table}
\end{center}

Table \ref{errorToyex2} shows that when $\beta$ is fixed and  $\alpha$ increases, resulting in an increase of the mean of the target variables at region level ($\alpha |\Omega|+ \beta x_\Omega$), all errors get larger. For fixed coefficients  $\alpha, \beta$, the errors increase from the first set of sources to the third one, which is natural since the available information decreases from  14 observations for the first case, to  4 for the third.

In accordance with the toy example of section \ref{firsttoy}, the errors for  $X_2$ are much smaller than the ones for $X_1$ for the areal weighting and the dasymetric methods due to the  difference of homogeneity of the auxiliary variables. The more homogeneous the auxiliary variable is, the more accurate the areal weighting interpolation and dasymetric methods are. As we discussed earlier, if the auxiliary information is almost homogeneous, the regression might be less exact than the areal weigthing. But Table 4 shows that the regression methods are still quite good for $X_2$: in general, they are better than the two classical methods except in some particular cases. The errors of the regression and scaled regression methods are very comparable for  $X_1$ and  $X_2$. Indeed, the prediction errors of the regression are very similar to the mean of the target variables $Y$ and the accuracy of the scaled regression predictor is equivalent to the one of the composite predictor.

The  effect of imbalance can be studied by looking at a change of $\alpha$ with fixed $\beta$. A larger $\alpha$ corresponds to a larger influence of the areal effect $\alpha |S|$ which is expected to lead to the domination of the areal weighting interpolation method (indeed we can observe this effect in the table for both  auxiliary variables $X_1$ and $ X_2$). The imbalance also affects the regression and scaled regression methods: if one of the two effects $\alpha |S|$ and $\beta X_S$ is much larger than the other one, the corresponding  errors seem to be further from their benchmarks (respectively the mean of $Y$ and the composite prediction error): see for example the cases $\alpha=1000, \beta=0.1$. However this effect is not very large  for regression and scaled regression. One factor which influences more these two methods is the homogeneity of the auxiliary variable: comparing  the results for $X_1$  and $X_2$ illustrates this. For $X_1$, the regression prediction errors are almost equal to their respective benchmarks and the amount of initial information does not seem to have a big influence (the errors are not monotonic from the first to the third set of sources). For $X_2,$  the accuracy increases with the number of source zones (the best being for the first one) and the errors of the regression method tend to the mean of $Y$. If we consider the particular case $\alpha=1000, \beta=0.1$ for  $X_2$, the areal impact is much stronger than the auxiliary information impact, and we see that the areal weighting interpolation is the best method, and that the scaled regression predictors can catch up the areal weighting interpolation when there are more source zones.

\noindent Table \ref{relativeErrorToyex2} contains the corresponding relative errors. For example, in the case $\alpha = 1000, \beta = 1$ for $X_2$, we see that whatever the number of sources the relative error is around $2.6\%$ to $2.7\%$ for the scaled regression (very close to the benchmark given by the last column) whereas dasymetric is around $4\%$ and areal weighting around $3\%$. Looking at the second column, we see that when the expected number of points increases, the relative prediction error tends to decrease which was naturally not the case for the prediction error itself.

\noindent We now turn attention to the robustness of the methods with respect to the model. As previously with the same geometrical design, we generate two auxiliary information scenarios: $X_1$ is as in the previous simulation, and $X_3$ is inhomogeneous and uncorrelated with $X_1$ (correlation coefficient of $-0.16$).
A target variable $Y$ is generated from $X_3$ with the relationship $Y_A \sim \P(600 |A| + X_3)$. We transfer $Y$ from the first set of 14 sources to the cells (Figure \ref{cellsandsources}) by using areal weighting  interpolation, dasymetric interpolation with $X_1$ and $X_3$ as auxiliary variables, the regression methods (REG and SCR) with the true model (areal effect and $X_3$), a simple model with only the areal effect, an  auxiliary variable model with an irrelevant variable (with area and $X_1$), an auxiliary variable model involving an unnecessary variable (the area and both $X_1$ and $ X_3$). Table \ref{tableRobust} presents the results.



\begin{table}[h]
\begin{center}
\begin{tabular}{|l|r|}
\hline
\multicolumn{1}{|c|}{ \bf Methods}  &  {\centering \textbf{Relative error}} \\
\hline
\hline
DAW&  7.74\\
\hline
\hline
DAX with $X_3$&9.49        \\
\hline
REG with area and $X_3$ &2.66 \\
\hline
ScR with area and $X_3$&2.62\\
\hline
\hline
DAX with $X_1$&14.70\\
\hline
REG with area and $X_1$&10.48\\
\hline
ScR with area and $X_1$&8.26 \\
\hline
\hline
REG with area &10.62\\
\hline
ScR with area&7.74\\
\hline
\hline
REG with area, $X_1$ and $X_3$& 2.66\\
\hline
ScR with area, $X_1$ and $X_3$&2.62\\
\hline
\end{tabular}
\end{center}
\caption{\bf Robusness of methods.}\label{tableRobust}
\end{table}

The most accurate method is the scaled regression with area and $X_3$ (true model). Note that the relative error for DAW and ScR with area only is the same which was expected since we proved that in that case the two methods coincide. The regression methods for the  model involving area plus $X_1$ and $X_3$ as auxiliary have the same errors (2.66\% and 2.62\%): in other words using unnecessary variables in the  regression does not decrease the accuracy. On the other hand, if we use the regression with a wrong choice of auxiliary variable, it gives bad predictions (10.48\% and 8.26\% for the model with area and $X_1$, 10.62\% and 7.74\% for the model with only areal effect). The dasymetric method with $X_3$ is better than with $X_1$ (9.49\% vs 14.70\%) which makes sense because the correlation of the target variable $Y$ with  $X_3$ is $0.998$ while with  $X_1$ it is of $-0.159$  however we see that despite the strong correlation  between $Y$ and $X_3$ the dasymetric method with $X_3$ is not so good because the areal effect is strong. The scaled regression is always better than the regression method and the scaled regression in the case of areal effect model yields the same result as the areal weighting interpolation method.

\section{Conclusion}
In this paper we have analyzed the accuracy of  four areal interpolation methods: areal weighting interpolation, dasymetric interpolation, Poisson regression and scaled Poisson regression for the case of count data. We have introduced a model based on an underlying Poisson point pattern to be able to evaluate the accuracy of the different methods.  We have proposed a scaled version of the Poisson regression method resulting in the enforcement of the  pycnophylactic property. Areal weighting interpolation and dasymetric interpolation have been compared with a finite distance approach and the regression methods have been compared together and with the previous ones with an  asymptotic approach.

We found out that one shouldn't rely on the correlation of the target variable and the auxiliary variable or on the homogeneity of the target variable to decide between areal interpolation or dasymetric but we should also take into account the relative imbalance between the areal effect and the auxiliary effect. A strong areal effect leads to the dominance of the areal weighting interpolation and a strong auxiliary effect is in favor of the dasymetric method. Moreover, the  imbalance index allows to approximate the ratio of the two relative errors and their lower bounds as the number of points on the source zones gets large. We establish the formula for the best linear predictor (therefore better than the areal weighting and the dasymetric), which leads to the introduction of the  scaled regression method.

For the comparison of areal weighting and dasymetric, a combination of several factors explains the complexity of the behavior: the size of sources, the auxiliary information, the number and size of target zones, \ldots  The error at source level
 is better when sources are divided into a smaller number of target zones. A large number of points makes the error at source level worse but improves the accuracy of the relative error. These two types of errors have the same behavior as a function of the imbalance index. 
 The impact of the expected number of points and of the inhomogeneity on the comparative advantage of the methods should not be forgotten: indeed when we have several sources, the sign of the imbalance index may vary from source to source  and the overall effect, being an aggregate of the source level effect, will also depend on the magnitude of the source error differences which is driven by the expected number of points and by the inhomogeneity. 
 We proved that the accuracy of the unfeasible composite predictor is decreasing when the expected number of points are similar on all targets and this fact extends to scaled regression (due to the approximation results).

To be able to include the regression methods in the comparison, we need to resort to some asymptotic approach. We propose an asymptotic framework and prove that the Poisson regression prediction error is equivalent to the variance of the underlying process and for the scaled regression, it is approximated by the composite's prediction error. These results show the regression predictor is not automatically better than the areal weighting interpolation or the dasymetric method, but when the number of points at source level is large, it is in general better. Finally the scaled regression turns out to be the best one among the considered methods. These results are confirmed by our simulation study of the last section.
The robustness with respect to the model is also considered.  The simulations show that a model with extra auxiliary variables doesn't create any loss while missing variables or unrelated variables (in place of the correct ones) decrease the accuracy of all methods.

\section{Appendix}\label{sectionAppendix}
\subsection{Proofs}
\subsubsection{Proof of Lemma \ref{bias} and lemma \ref{variances}}
From \eqref{DAW}, \eqref{DAX} and the properties of a Poisson point process we have
 \begin{align*}
  \E(\hat{Y}^{DAW}_{T}-Y_T)&= \E(\dfrac{|T|}{|S|} Y_S-Y_T)=\dfrac{|T|}{|S|}(\alpha |S|+\beta x_S)-(\alpha |T|+\beta x_T)=\beta x_S (\dfrac{|T|}{|S|}-\dfrac{x_T}{x_S})
  \end{align*}
  \begin{align*}
  \E(\hat{Y}^{DAX}_{T}-Y_T)&= \E(\dfrac{x_T}{x_S} Y_S-Y_T)=\dfrac{x_T}{x_S}(\alpha |S|+\beta x_S)-(\alpha |T|+\beta x_T)=\alpha |S| (\dfrac{x_T}{x_S}-\dfrac{|T|}{|S|})
  \end{align*}

  Taking into account the independence of two disjoint target zones with the fact that the target $T$ is a portion of the source $S$ the variances of each method are given as follows
  \begin{align*}
  Var(\hat{Y}^{DAW}_{T}-Y_T)&=Var(\dfrac{|T|}{|S|}Y_S-Y_T)\\
  &=\dfrac{|T|^2}{|S|^2} Var(Y_S)+Var(Y_T)-2\dfrac{|T|}{|S|}Cov(Y_S, Y_T)\\
  &=\dfrac{|T|^2}{|S|^2} \E(Y_S)+\E(Y_T)-2\dfrac{|T|}{|S|}Cov(Y_{S\setminus T}+Y_T, Y_T)\\
  &=\dfrac{|T|^2}{|S|^2} \E(Y_S)+\E(Y_T)-2\dfrac{|T|}{|S|}Var(Y_T)\\
  &=\dfrac{|T|^2}{|S|^2} (\alpha |S|+\beta x_S)+(\alpha |T|+\beta x_T)-2\dfrac{|T|}{|S|}(\alpha |T|+\beta x_T)\\
  &=\beta x_S (\dfrac{|T|}{|S|}-\dfrac{x_T}{x_S})^2+\beta x_T(1-\dfrac{x_T}{x_S}) +\alpha |T| (1-\dfrac{|T|}{|S|})
    \end{align*}
      \begin{align*}
  Var(\hat{Y}^{DAX}_{T}-Y_T)&=Var(\dfrac{x_T}{x_S}Y_S-Y_T)\\
  &=\dfrac{x_T^2}{x_S^2} Var(Y_S)+Var(Y_T)-2\dfrac{x_T}{x_S}Cov(Y_S, Y_T)\\
  &=\dfrac{x_T^2}{x_S^2} \E(Y_S)+\E(Y_T)-2\dfrac{x_T}{x_S}Cov(Y_{S\setminus T}+Y_T, Y_T)\\
  &=\dfrac{x_T^2}{x_S^2} \E(Y_S)+\E(Y_T)-2\dfrac{x_T}{x_S}Var(Y_T)\\
  &=\dfrac{x_T^2}{x_S^2} (\alpha |S|+\beta x_S)+(\alpha |T|+\beta x_T)-2\dfrac{x_T}{x_S}(\alpha |T|+\beta x_T)\\
  &=\alpha |S| (\dfrac{|T|}{|S|}-\dfrac{x_T}{x_S})^2+\beta x_T(1-\dfrac{x_T}{x_S}) +\alpha |T| (1-\dfrac{|T|}{|S|})
    \end{align*}

 Summing up the variances at target level with the fact that $\sum_T {|T| \over |S|}=\sum_T {x_T\over x_S}=1$, we get the variances at source level
    \begin{align*}
    Var_S^{DAW}&=\sum_T Var(\hat{Y}^{DAW}_{T}-Y_T)\\
    &=\sum_T \beta x_S (\dfrac{|T|}{|S|}-\dfrac{x_T}{x_S})^2+\beta x_T(1-\dfrac{x_T}{x_S}) +\alpha |T| (1-\dfrac{|T|}{|S|})\\
    &=\beta x_S \sum_T(\dfrac{|T|}{|S|}-\dfrac{x_T}{x_S})^2+\beta x_S(1-\sum_T \dfrac{x_T^2}{x_S^2}) +\alpha |S| (1-\sum_T\dfrac{|T|^2}{|S|^2})
    \end{align*}

    \begin{align*}
    Var_S^{DAX}&=\sum_T Var(\hat{Y}^{DAX}_{T}-Y_T)\\
 &=\alpha |S| \sum_T(\dfrac{|T|}{|S|}-\dfrac{x_T}{x_S})^2+\beta x_S(1-\sum_T \dfrac{x_T^2}{x_S^2}) +\alpha |S| (1-\sum_T\dfrac{|T|^2}{|S|^2})
    \end{align*}


 \subsubsection{Proof of Theorem \ref{reler}}
 From Lemma \ref{bias} and the fact that $\alpha |S|=I_S(|.|)\E(Y_S), \beta X_S=I_S(X)\E(Y_S)$ we have
 \begin{align*}
 \Er_T^{DAW}&= I_S(X)\E(Y_S) (\dfrac{|T|}{|S|}-\dfrac{x_T}{x_S})^2+I_S(X)\E(Y_S)(\dfrac{x_T}{x_S} - \dfrac{x_T^2}{x_S^2}) +I_S(|.|)\E(Y_S) (\dfrac{|T|}{|S|} -\dfrac{|T|^2}{|S|^2})\\
 &+ I_S(X)^2\E(Y_S)^2 (\dfrac{|T|}{|S|}-\dfrac{x_T}{x_S})^2\\
\Er_T^{DAX}&= I_S(|.|)\E(Y_S) (\dfrac{|T|}{|S|}-\dfrac{x_T}{x_S})^2+I_S(X)\E(Y_S)(\dfrac{x_T}{x_S} - \dfrac{x_T^2}{x_S^2}) +I_S(|.|)\E(Y_S) (\dfrac{|T|}{|S|} -\dfrac{|T|^2}{|S|^2})\\
&+ I_S(|.|)^2\E(Y_S)^2 (\dfrac{|T|}{|S|}-\dfrac{x_T}{x_S})^2
 \end{align*}
  If the expectation of the number of points is sufficiently large, we can approximate the ratio of the two errors as follows
  $$\dfrac{\Er_T^{DAW}}{\Er_T^{DAX}}\approx\dfrac{I_S(X)^2}{I_S(|.|)^2}$$
 and also
$$\dfrac{\Re_T^{DAW}}{\Re_T^{DAX}}\approx\dfrac{I_S(X)}{I_S(|.|)}$$
At source level, we get a similar result by adding up errors on all target zones using the fact that $\sum_{T}\dfrac{|T|}{|S|}=\sum_{T}\dfrac{x_T}{x_S}=1$

  \begin{align*}
 \Er_S^{DAW}&= I_S(X)\E(Y_S)\sum_{T} (\dfrac{|T|}{|S|}-\dfrac{x_T}{x_S})^2+I_S(X)\E(Y_S)(1 -\sum_{T} \dfrac{x_T^2}{x_S^2}) +I_S(|.|)\E(Y_S) (1 -\sum_{T}\dfrac{|T|^2}{|S|^2})\\
 &+ I_S(X)^2\E(Y_S)^2\sum_{T} (\dfrac{|T|}{|S|}-\dfrac{x_T}{x_S})^2\\
\Er_S^{DAX}&= I_S(|.|)\E(Y_S) (1-\sum_{T}\dfrac{x_T}{x_S})^2+I_S(X)\E(Y_S)(1 - \sum_{T}\dfrac{x_T^2}{x_S^2}) +I_S(|.|)\E(Y_S) (1 -\sum_{T}\dfrac{|T|^2}{|S|^2})\\
&+ I_S(|.|)^2\E(Y_S)^2 \sum_{T}(\dfrac{|T|}{|S|}-\dfrac{x_T}{x_S})^2\\
&\Rightarrow\\
\Re_S^{DAW}&= \dfrac{1}{\E(Y_S)} [I_S(X)\sum_{T} (\dfrac{|T|}{|S|}-\dfrac{x_T}{x_S})^2+I_S(X)(1 -\sum_{T} \dfrac{x_T^2}{x_S^2}) +I_S(|.|) (1 -\sum_{T}\dfrac{|T|^2}{|S|^2})]\\
 &+ I_S(X)^2\sum_{T} (\dfrac{|T|}{|S|}-\dfrac{x_T}{x_S})^2\\
\Re_S^{DAX}&= \dfrac{1}{\E(Y_S)} [I_S(|.|) (1-\sum_{T}\dfrac{x_T}{x_S})^2+I_S(X)(1 - \sum_{T}\dfrac{x_T^2}{x_S^2}) +I_S(|.|) (1 -\sum_{T}\dfrac{|T|^2}{|S|^2})]\\
&+ I_S(|.|)^2 \sum_{T}(\dfrac{|T|}{|S|}-\dfrac{x_T}{x_S})^2
 \end{align*}
Using the relationship $I_S(|.|)+I_S(X)=1$, the above results prove Theorem \ref{reler}.

 \subsubsection{Proof of Theorem \ref{TheoDiffErDaxDaw}}
Lemma \ref{bias} yields
 \begin{align*}
  \Er^{DAW}_T&-\Er^{DAX}_T=Var(\hat{Y}^{DAW}_{T}-Y_T)+[\E(\hat{Y}^{DAW}_{T}-Y_T)]^2  \\
  &-Var(\hat{Y}^{DAX}_{T}-Y_T)-[\E(\hat{Y}^{DAX}_{T}-Y_T)]^2\\
  &= (\dfrac{|T|}{|S|}-\dfrac{x_T}{x_S})^2(\beta x_S-\alpha |S|) +(\dfrac{|T|}{|S|}-\dfrac{x_T}{x_S})^2(\beta^2 x_S^2-\alpha^2 |S|^2)\\
  &=(\dfrac{|T|}{|S|}-\dfrac{x_T}{x_S})^2(\beta x_S-\alpha |S|)((\beta x_S+\alpha |S|+1)\\
  &=(\dfrac{|T|}{|S|}-\dfrac{x_T}{x_S})^2\dfrac{(\beta x_S-\alpha |S|)}{(\beta x_S+\alpha |S|)}((\beta x_S+\alpha |S|+1)(\beta x_S+\alpha |S|)\\
  &=(\dfrac{|T|}{|S|}-\dfrac{x_T}{x_S})^2 \Delta_S (\E(Y_S)+1)\E(Y_S)
  \end{align*}


 \subsubsection{Proof of Theorem \ref{composite}}
  We calculate the error of the composite predictors then minimize with respect to $w$ to find the optimal $w^*$

\begin{align*}
\hat{Y}^C_T&= w\hat{Y}^{DAW}_{T}+(1-w)\hat{Y}^{DAX}_{T}=[w \dfrac{|T|}{|S|}+ (1-w)\dfrac{x_T}{x_S}]Y_S := u Y_S\\
Bias_T^2&=[\E(\hat{Y}^{DAW}_{T}-\hat{Y}^{DAX}_{T})]^2=(u\lambda_S -\lambda_T)^2\\
Var_T&=Var(u Y_S-Y_T)=u^2 \lambda_S + \lambda_T -2 u\lambda_T\\
\Er_T&=u^2\lambda_S(\lambda_S+1) -2u\lambda_T(\lambda_S+1)+\lambda_T^2+\lambda_T\\
u^*&=argmin_u  \Er_T=\dfrac{\lambda_T}{\lambda_S}\\
\Leftrightarrow & w^* \dfrac{|T|}{|S|}+ (1-w^*)\dfrac{x_T}{x_S}=\dfrac{\alpha |T|+\beta x_T}{\alpha |S|+\beta x_S}\\
\Leftrightarrow & w^*=\dfrac{\alpha |T|}{\alpha |S|+\beta x_S}
\end{align*}
Substituting the $w^*$ in \eqref{compositeclass} we get the composite predictor \eqref{Com}.

The bias, variance and error of the above composite predictor are calculated as follows
 \begin{align*}
Bias=\E(\hat{Y}^C_T-Y_T)&=0\\
\Er^C_T=Var(\hat{Y}^C_T-Y_T)&=Var(\dfrac{\lambda_T}{\lambda_S}Y_S -Y_T)\\
&=\dfrac{\lambda_T^2}{\lambda_S^2}Var(Y_S) + Var(Y_T)-2\dfrac{\lambda_T}{\lambda_S} Cov(Y_S,Y_T)\\
&=\dfrac{\lambda_T^2}{\lambda_S^2} \lambda_S +\lambda_T-2 \dfrac{\lambda_T}{\lambda_S} \lambda_T\\
&=\lambda_T - \dfrac{\lambda_T^2}{\lambda_S}\\
&=\dfrac{x_T^2}{x_S^2} \lambda_S+\lambda_T-2\dfrac{x_T}{x_S}\lambda_T -\dfrac{x_T^2}{x_S^2} \lambda_S+2\dfrac{x_T}{x_S}\lambda_T- \dfrac{\lambda_T^2}{\lambda_S}\\
&=\dfrac{x_T^2}{x_S^2} \lambda_S+\lambda_T-2\dfrac{x_T}{x_S}\lambda_T -\lambda_S(\dfrac{x_T}{x_S}-\dfrac{\lambda_T}{\lambda_S})^2\\
&=Var(\hat{Y}^{DAX}_T-Y_T)-\lambda_S(\dfrac{x_T}{x_S}-\dfrac{\lambda_T}{\lambda_S})^2\\
&=Var(\hat{Y}^{DAW}_T-Y_T)-\lambda_S(\dfrac{|T|}{|S|}-\dfrac{\lambda_T}{\lambda_S})^2
\end{align*}

Since
\begin{align*}
Y_T|Y_S \sim Bi(Y_S, \dfrac{\E(Y_T)}{\E(Y_S)})
\end{align*}
we have
\begin{align*}
\E(Y_T|Y_S)= \dfrac{\E(Y_T)}{\E(Y_S)}Y_S=\hat{Y}^C_T
\end{align*}
This shows that the composite predictor is the best linear predictor.

\subsubsection{Proof of Theorem \ref{ConsitencyNormality}}
To prove the theorem, we will prove the following lemmas

\begin{lemma}\label{lemNormedScore}
Under conditions (C1) and (C2), the normed score function $F_n^{-1/2}s_n$ is asymptotically normal
\begin{equation}
F_n^{-1/2}s_n \to_d \N(0,\I)
\end{equation}
 \end{lemma}
 \begin{lemma}\label{normed2derivative}
 Under conditions (C1) and (C2), for all $\delta >0$
 \begin{equation}\label{EqNormed2Deri}
 max_{\gamma \in N_n(\delta)} ||V_n(\gamma)-\I|| \to_p 0
 \end{equation}
where $N_n(\delta)=\{\gamma: ||F_n^{1/2} (\gamma - \gamma_o)||\leq \delta\}, V_n (\gamma)=F_n^{-1/2}H_n(\gamma)F_n^{-1/2}$.
 \end{lemma}

Lemma \ref{lemNormedScore} is proved by using the Lindeberg-Feller theorem.

Indeed, for  $\tau$ fixed with $\tau' \tau =1$, considering the triangular array
\begin{equation}
z_{n,i}=\tau' F_n^{-1/2} \dfrac{\tilde{Z}_{n,i}}{\gamma' \tilde{Z}_{n,i}}(y_{n,i}-\gamma' \tilde{Z}_{n,i})
\end{equation}
we have
\begin{align*}
\E(z_{n,i})&=0\\
\sum_i Var(z_{n,i})&=1
\end{align*}
We will show that the Lindeberg condition is satisfied, i.e. for any $\varepsilon >0$
\begin{equation}\label{LinCondition}
\sum_i  \E(z_{n,i}^2 \1_{|z_{n,i}|>\varepsilon}) \to 0
\end{equation}
   as $n\to \infty$.

Let $a_{n,i}=\tau' F_n^{-1/2} \dfrac{\tilde{Z}_{n,i}}{\gamma' \tilde{Z}_{n,i}}$, because $z_{n,i}^2  = a_{n,i}^2 (y_{n,i}-\gamma' \tilde{Z}_{n,i})^2$ ,  $\E(z_{n,i}^2 \1_{|z_{n,i}|>\varepsilon}) = a_{n,i}^2\E((y_{n,i}-\gamma' \tilde{Z}_{n,i})^2\1_{|y_{n,i}-\gamma' \tilde{Z}_{n,i}|>\frac{\varepsilon}{|a_{n,i}|}})$ , yields

\begin{align*}
\sum_i  \E(z_{n,i}^2 \1_{|z_{n,i}|>\varepsilon}) &=\sum_i a_{n,i}^2\E((y_{n,i}-\gamma' \tilde{Z}_{n,i})^2\1_{|y_{n,i}-\gamma' \tilde{Z}_{n,i}|>\frac{\varepsilon}{|a_{n,i}|}})\\
&\leq (\sum_i a_{n,i}^2 )sup_{i}\E((y_{n,i}-\gamma' \tilde{Z}_{n,i})^2\1_{|y_{n,i}-\gamma' \tilde{Z}_{n,i}|>\frac{\varepsilon}{|a_{n,i}|}})
\end{align*}

Moreover, condition (C1) yields that there is a positive number $K_1$ s.t. ${1\over \gamma' \tilde{Z}_{n,i}}<K_1, \forall (n,i)$, hence
\begin{align*}
\sum_i a_{n,i}^2 &=\tau' F_n^{-1/2}\sum_i \dfrac{\tilde{Z}_{n,i}\tilde{Z}'_{n,i}}{(\gamma' \tilde{Z}_{n,i})^2} F_n^{-1/2}\tau\\
&<K_1 \tau' F_n^{-1/2}\sum_i \dfrac{\tilde{Z}_{n,i}\tilde{Z}'_{n,i}}{\gamma' \tilde{Z}_{n,i}} F_n^{-1/2}\tau=K_1
\end{align*}
In addition, conditions (C1) (C2) lead to
$$max_i \dfrac{\varepsilon}{|a_{n,i}|} \to \infty$$
as $n \to \infty$, hence for any $M>0, \exists n_1$ s.t. $\forall n>n_1$
\begin{align*}
sup_{i}&\E((y_{n,i}-\gamma' \tilde{Z}_{n,i})^2\1_{|y_{n,i}-\gamma' \tilde{Z}_{n,i}|>\frac{\varepsilon}{|a_{n,i}|}})\leq sup_{i}\E((y_{n,i}-\gamma' \tilde{Z}_{n,i})^2\1_{|y_{n,i}-\gamma' \tilde{Z}_{n,i}|>M})\\
&\leq sup_{i}\sqrt{\E((y_{n,i}-\gamma' \tilde{Z}_{n,i})^4\E(\1_{|y_{n,i}-\gamma' \tilde{Z}_{n,i}|>M})}\\
&\leq sup_{i}\sqrt{\E((y_{n,i}-\gamma' \tilde{Z}_{n,i})^4\dfrac{Var(y_{n,i}-\gamma' \tilde{Z}_{n,i})}{M^2}}\\
&=sup_{i}\sqrt{\gamma' \tilde{Z}_{n,i}(1+3\gamma' \tilde{Z}_{n,i})\dfrac{\gamma' \tilde{Z}_{n,i}}{M^2}}\\
&< {K_2 \over M}
\end{align*}
hence
$$sup_{i}\E((y_{n,i}-\gamma' \tilde{Z}_{n,i})^2\1_{|y_{n,i}-\gamma' \tilde{Z}_{n,i}|>\frac{\varepsilon}{|a_{n,i}|}}) \to 0 \text{ as } n \to \infty $$
where the existence of $K_2$ is derived from condition (C1). This argument shows that the \eqref{LinCondition} holds. So does Lemma \ref{lemNormedScore}.

\noindent {\it Proof of Lemma \ref{normed2derivative}}

Using the same notation in the proof of Lemma \ref{lemNormedScore}, $\tau$ fixed s.t. $\tau'\tau=1$, let $b_{n,i}=\tau' F_n^{-1/2} \tilde{Z}_{n,i}$, the equation \eqref{EqNormed2Deri} can be rewritten as
\begin{equation}
\tau'(V_n (\gamma)-\I)\tau=A_n+B_n+C_n
\end{equation}
where
\begin{align}
A_n&=\sum_i b_{n,i}^2 ({1 \over (\gamma' \tilde{Z}_{n,i})^2}-{1 \over (\gamma'_o \tilde{Z}_{n,i})^2})(y_{n,i}-\gamma' \tilde{Z}_{n,i})\label{An}\\
B_n&=\sum_i b_{n,i}^2 {1 \over (\gamma'_o \tilde{Z}_{n,i})^2}(y_{n,i}-\gamma' \tilde{Z}_{n,i})\label{Bn}\\
C_n&=\sum_i b_{n,i}^2 ({1 \over \gamma' \tilde{Z}_{n,i}}-{1 \over \gamma'_o \tilde{Z}_{n,i}})\label{Cn}
\end{align}

We will prove that the three terms converge in probability to 0 as $n$ tends to  $\infty$.
To prove \eqref{Bn}, we first study its properties. We have
\begin{align*}
\E(B_n)&=0\\
Var(B_n)&=\sum_i b_{n,i}^4 {1 \over (\gamma'_o \tilde{Z}_{n,i})^4}Var(y_{n,i}-\gamma' \tilde{Z}_{n,i})\\
&=\sum_i b_{n,i}^4 {1 \over (\gamma'_o \tilde{Z}_{n,i})^4}\gamma' \tilde{Z}_{n,i}\\
&\leq \sum_i b_{n,i}^2 {1 \over \gamma'_o \tilde{Z}_{n,i}}\sup_i {b_{n,i}^2 \over (\gamma'_o \tilde{Z}_{n,i})^3}\gamma' \tilde{Z}_{n,i}\\
&=\sup_i b_{n,i}^2 {1 \over (\gamma'_o \tilde{Z}_{n,i})^3}\gamma' \tilde{Z}_{n,i}<K_3 \sup_i b_{n,i}^2
\end{align*}
Because of the boundedness of $(\gamma'_o \tilde{Z}_{n,i})^3$ and the definition of $N_n(\delta)$, $\gamma' \tilde{Z}_{n,i}$ is bounded when $n$ is large enough, moreover, $\sup_i b_{n,i}^2 \to 0$ due to the condition (C1) (C2), therefore
 $$B_n \to_p 0$$

We can use similar argument to prove $A_n \to_p 0, C_n \to 0$, and this shows that the lemma \ref{normed2derivative} holds.

\subsubsection{Proof of Theorem \ref{TheNormalityREG}}


Let $$z_{nij}\sim \P(\gamma_o'{Z}_{n,i})-\gamma_o'{Z}_{n,i}:=\tilde{z}_{n,i},\, j=1,2,...,k_n \text{ i.i.d }$$

This yields
$$\sum_j z_{nij} = Y_{n,i}-\gamma_o'\tilde{Z}_{n,i}$$
We have

\begin{align*}
& \E(z_{nij})=0\\
& \sum_j Var(z_{nij})=\gamma_o'\tilde{Z}_{n,i}
\end{align*}

We will prove that this array satisfies the Lindeberg-Feller condition, i.e. $\forall \delta >0$

$$\sum_j \E(z_{nij}^2 \1_{|z_{nij}|> \delta})\to 0, \text{ as } n \to \infty $$

Indeed,
\begin{align*}
\sum_j \E(z_{nij}^2 \1_{|z_{nij}|> \delta})=&k_n \E(\tilde{z}_{n,i}^2 \1_{|\tilde{z}_{n,i}|> \delta})=\E(u_{n,i}^2 \1_{|u_{n,i}|> \sqrt{k_n} \delta})
\end{align*}
 where $u_{n,i}=\sqrt{k_n} \tilde{z}_{n,i}$. Because $\E u_{n,i}=0$, $\E u_{n,i}^2=Var u_{n,i}=k_n Var \tilde{z}_{n,i}=\gamma_o'\tilde{Z}_{n,i} < \infty$. Moreover $k_n \to \infty$ as $n \to \infty$, we have
 $$\E(u_{n,i}^2 \1_{|u_{n,i}|> \sqrt{k_n} \delta}) \to 0$$
 as $n \to \infty$

 From the  Lindeberg-Feller theorem we get

 $${Y_{n,i} - \gamma_o'\tilde{Z}_{n,i} \over  \sqrt{\gamma_o'\tilde{Z}_{n,i}}}  \to_d \N(0, 1)$$

This proof can be applied at the target level, i.e.
$${Y_{T} - \gamma_o' \tilde{Z}_{T} \over  \sqrt{\gamma_o'\tilde{Z}_{T}}}  \to_d \N(0, 1)$$

\subsubsection{Proof of Proposition \ref{ProPycno}}
The pycnophylactic property of the scaled regression predictor is obvious.

To prove the pycnophylactic property of the regression predictor at region level, we sum up regression predictors over source zones
\begin{align*}
\hat{Y}_{\Omega}=\sum_i\sum_{T: T \subset S_{n,i}}\hat{Y}^{REG}_{T}&=\sum_i\sum_{T: T \subset S_{n,i}} \hat{\gamma}_n\tilde{Z}_{T}=\hat{\gamma}_n'\tilde{Z}_\Omega
\end{align*}

Recall that  $\hat{\gamma}$ is the solution of the score equation $s_n(\gamma)=0$, i.e.
\begin{align*}
&\sum_{i=1}^n \dfrac{\tilde{Z}_{n,i}}{\hat{\gamma}' \tilde{Z}_{n,i}} y_{n,i} - \tilde{Z}_{n,i}=0\\
\Rightarrow &\sum_{i=1}^n \dfrac{\hat{\gamma}'\tilde{Z}_{n,i}}{\hat{\gamma}' \tilde{Z}_{n,i}} y_{n,i} - \hat{\gamma}'\tilde{Z}_{n,i}=0\\
\Leftrightarrow &\sum_{i=1}^n  y_{n,i} - \hat{\gamma}'\tilde{Z}_{\Omega}=0\\
\Leftrightarrow &\hat{\gamma}'\tilde{Z}_{\Omega}=y_{\Omega}
\end{align*}
In other words, the regression predictor satisfies the pycnophylactic property on the region $\Omega$.

To study the pycnophylactic property of the regression predictor at source level, we consider
 $$\hat{Y}_{n,i}^{REG}-Y_{n,i}$$
We have
\begin{align*}
\hat{Y}_{n,i}^{REG}-Y_{n,i}&=\hat{\gamma}_n'\tilde{Z}_{n,i}-Y_{n,i}\\
&=(\hat{\gamma}_n'-\gamma_o' )\tilde{Z}_{n,i}-(Y_{n,i}-\gamma_o' \tilde{Z}_{n,i})\\
&=F_n^{-1/2}F_n^{1/2}(\hat{\gamma}_n'-\gamma_o' )\tilde{Z}_{n,i}-(Y_{n,i}-\gamma_o'\tilde{Z}_{n,i})\\
\end{align*}
The first term converges to 0 in distribution due to the conditions (C1), (C2) and the theorem \ref{ConsitencyNormality}. The second term is different from 0, even asymptotically (Proposition \ref{ProPycno}).

Moreover, because of the boundedness of $\tilde{Z}_{n,i}$, the above argument yields
$$\dfrac{\hat{Y}_{n,i}^{REG}-Y_{n,i}}{\sqrt{\gamma_o'\tilde{Z}_{n,i}}} \to_d \N(0,1)$$

This completes the proof of proposition \ref{ProPycno}.

If $\tilde{Z}_{T}$ is  bounded below, a similar result at target level holds
$$\dfrac{\hat{Y}_{T}^{REG}-Y_{T}}{\sqrt{\gamma_o'\tilde{Z}_{T}}} \to_d \N(0,1)$$

\subsubsection{Proof of Theorem \ref{TheErrReg}}
For any target $T$, the error of the regression predictor on the target is

\begin{align*}
\E(\hat{Y}^{REG}_T-Y_T)^2&=\E( \hat{\gamma}_n' \tilde{Z}_T-{\gamma}_o'\tilde{Z}_T )^2+\E({\gamma}_o'\tilde{Z}_T -Y_T)^2-2\E( \hat{\gamma}_n' \tilde{Z}_T-{\gamma}_o'\tilde{Z}_T )({\gamma}_o'\tilde{Z}_T -Y_T)
\end{align*}

From Theorem \ref{ConsitencyNormality} and condition (C1), for any $\eta_1>0, \exists \varepsilon >0$ s.t.when $n$ is sufficiently large
\begin{align}
\E(\hat{\gamma}_n'\tilde{Z}_T -{\gamma}_o'\tilde{Z}_T )^2 \1_{||\hat{\gamma}_n-{\gamma}_o||< \varepsilon}&<\eta_1\label{eq5.1.1}\\
||2\E( \hat{\gamma}_n' \tilde{Z}_T-{\gamma}_o'\tilde{Z}_T )({\gamma}_o'\tilde{Z}_T -Y_T)\1_{||\hat{\gamma}_n-{\gamma}_o||< \varepsilon}||&<\eta_1\label{eq5.1.2}
\end{align}
As we proved in  Proposition \ref{ProPycno} $( \hat{\gamma}_n' \tilde{Z}_T-{\gamma}_o'\tilde{Z}_T ) \to_p 0$, we have
\begin{align*}
\Pr(||\hat{\gamma}_n-{\gamma}_o||< \varepsilon)&\to 1 \text{ as } n \to \infty
\end{align*}
In addition
$$\E({\gamma}_o'\tilde{Z}_T -Y_T)^2={\gamma}_o'\tilde{Z}_T $$
Hence there is $n_1$ s.t.
$$\E({\gamma}_o'\tilde{Z}_T -Y_T)^2\1_{||\hat{\gamma}_n-{\gamma}_o||\geq \varepsilon}<
\E({\gamma}_o'\tilde{Z}_T -Y_T)^4 \Pr(||\hat{\gamma}_n-{\gamma}_o||\geq \varepsilon)<\eta_1$$
for $n>n_1$. In other words,
$${\gamma}_o'\tilde{Z}_T  -\eta_1<\E({\gamma}_o'\tilde{Z}_T -Y_T)^2\1_{||\hat{\gamma}_n-{\gamma}_o||> \varepsilon}<{\gamma}_o'\tilde{Z}_T $$

This implies $\forall \eta >0, \exists \varepsilon >0, n_1$ s.t. for $n>n_1$
\begin{align}
-\eta +{\gamma}_o'\tilde{Z}_T <\E(\hat{Y}^R_T-Y_T)^2\1_{||\hat{\gamma}_n-{\gamma}_o||< \varepsilon}<\eta +{\gamma}_o'\tilde{Z}_T \label{eq5.1.3}
\end{align}
with a remark that $\Pr(||\hat{\gamma}_n-{\gamma}_o||< \varepsilon)\to 1 \text{ as } n \to \infty$.

Combining \eqref{eq5.1.1}, \eqref{eq5.1.2}, \eqref{eq5.1.3} we get Theorem \ref{TheErrReg}.
\subsubsection{Proof of equations (\ref{diffapproRelEr})}
We rewrite the error of the areal interpolation and dasymetric for the asymptotic model. For a target $T\subset S_{n,i}$, from \eqref{ErComDAX}, \eqref{ErComDAW}, and Lemma \ref{bias} we have
\begin{align*}
\Er^{DAW}_T &= \beta^2 \tilde{x}_{n,i}
\gamma_o'\tilde{Z}_T-\dfrac{(\gamma_o'\tilde{Z}_T)^2}{\gamma_o'\tilde{Z}_{n,i}}+\gamma_o'\tilde{Z}_{n,i}(\dfrac{\widetilde{|T|}}{\widetilde{|S_{n,i}|}}-\dfrac{\gamma_o'\tilde{Z}_T}{\gamma_o'\tilde{Z}_{n,i}})^2+\beta^2 \tilde{x}_S^2(\dfrac{\widetilde{|T|}}{\widetilde{|S_{n,i}|}}-\dfrac{\tilde{x}_T}{\tilde{x}_{n,i}})^2\\
\Er^{DAX}_T &=\gamma_o'\tilde{Z}_T-\dfrac{(\gamma_o'\tilde{Z}_T)^2}{\gamma_o'\tilde{Z}_{n,i}}+\gamma_o'\tilde{Z}_{n,i}(\dfrac{\tilde{x}_T}{\tilde{x}_{n,i}}-\dfrac{\gamma_o'\tilde{Z}_T}{\gamma_o'\tilde{Z}_{n,i}})^2+\alpha^2 \widetilde{|S_{n,i}|}^2(\dfrac{\widetilde{|T|}}{\widetilde{|S_{n,i}|}}-\dfrac{\tilde{x}_T}{\tilde{x}_{n,i}})^2
\end{align*}
A similar argument as in the proof of theorem \ref{TheErrReg} shows that,  for any $\eta_1>0, \varepsilon>0$, $\exists n_1$ s.t. $\forall n>n_1$ \begin{align*}
\Er^{DAW}_T-\eta_1&<\E(\hat{Y}_{T}^{DAW}-Y_{T})^2\1_{||\hat{\gamma}_n-{\gamma}_o||<\varepsilon}<\Er^{DAW}_T\\
\Er^{DAX}_T-\eta_1&<\E(\hat{Y}_{T}^{DAX}-Y_{T})^2\1_{||\hat{\gamma}_n-{\gamma}_o||<\varepsilon}<\Er^{DAX}_T
\end{align*}
With $\varepsilon$ chosen as in theorem \ref{TheErrReg}, let $Q_i=\{||\hat{\gamma}_n-{\gamma}_o||<\varepsilon\}$, we have
\begin{align*}
{\gamma}_o'\tilde{Z}_T-\Er^{DAW}_T -\eta<\E(\hat{Y}_{T}^{REG}-Y_{T})^2\1_{Q_i}-\E(\hat{Y}_{T}^{DAW}-Y_{T})^2\1_{Q_i} & < {\gamma}_o'\tilde{Z}_T -\Er^{DAW}_T+\eta+\eta_1\\
{\gamma}_o'\tilde{Z}_T -\Er^{DAX}_T-\eta<\E(\hat{Y}_{T}^{REG}-Y_{T})^2\1_{Q_i}-\E(\hat{Y}_{T}^{DAX}-Y_{T})^2\1_{Q_i} & < {\gamma}_o'\tilde{Z}_T -\Er^{DAX}_T+\eta+\eta_1
\end{align*}
for all $n>n_1$. Moreover,
\begin{align*}
{\gamma}_o'\tilde{Z}_T -\Er^{DAW}_T
&=\beta \tilde{x}_{n,i} {\tilde{x}_T^2 \over \tilde{x}_{n,i}^2} +\alpha \widetilde{|S_{n,i}|} {\widetilde{|T|}^2 \over \widetilde{|S_{n,i}|}^2}-\beta \tilde{x}_{n,i}(\dfrac{\widetilde{|T|}}{\widetilde{|S_{n,i}|}}-\dfrac{\tilde{x}_T}{\tilde{x}_{n,i}})^2-\beta^2 \tilde{x}_{n,i}^2(\dfrac{\widetilde{|T|}}{\widetilde{|S_{n,i}|}}-\dfrac{\tilde{x}_T}{\tilde{x}_{n,i}})^2\\
{\gamma}_o'\tilde{Z}_T -\Er^{DAW}_T
&=\beta \tilde{x}_{n,i} {\tilde{x}_T^2 \over \tilde{x}_{n,i}^2} +\alpha \widetilde{|S_{n,i}|} {\widetilde{|T|}^2 \over \widetilde{|S_{n,i}|}^2}-\alpha \widetilde{|S_{n,i}|}(\dfrac{\widetilde{|T|}}{\widetilde{|S_{n,i}|}}-\dfrac{\tilde{x}_T}{\tilde{x}_{n,i}})^2-\alpha^2 \widetilde{|S_{n,i}|}^2(\dfrac{\widetilde{|T|}}{\widetilde{|S_{n,i}|}}-\dfrac{\tilde{x}_T}{\tilde{x}_{n,i}})^2
\end{align*}
Taking the sum over all target zones which belong to $S_{n,i}$ then scaling the sum by $\E(Y_{n,i})$ and calculating the differences in terms of $\Delta_{n,i}=\Delta_{S_{n,i}}$, we have

\begin{align*}
4{\sum_T {\gamma}_o'\tilde{Z}_T -\Er^{DAW}_T \over \E(Y_{n,i})^2}&=4{1 \over \E(Y_{n,i})}[{\beta \tilde{x}_{n,i} \over \E(Y_{n,i})}\sum_T {\tilde{x}_T^2 \over \tilde{x}_{n,i}^2} +{\alpha \widetilde{|S_{n,i}|}\over \E(Y_{n,i})}\sum_T {\widetilde{|T|}^2  \over \widetilde{|S_{n,i}|}^2}-{\beta \tilde{x}_{n,i} \over \E(Y_{n,i})}\sum_T(\dfrac{\widetilde{|T|}}{\widetilde{|S_{n,i}|}}-\dfrac{\tilde{x}_T}{\tilde{x}_{n,i}})^2]\\
&-4({\beta\tilde{x}_{n,i}\over \E(Y_{n,i})})^2 \sum_T (\dfrac{\widetilde{|T|}}{\widetilde{|S_{n,i}|}}-\dfrac{\tilde{x}_T}{\tilde{x}_{n,i}})^2\\
&=2{1 \over \E(Y_{n,i})}[(1 + \Delta_{n,i})\sum_T {{x}_T^2 \over {x}_S^2} +(1 - \Delta_{n,i})\sum_T {|T|^2  \over |S_{n,i}|^2}-\\
&-(1 + \Delta_{n,i})\sum_T(\dfrac{{|T|}}{{|S_{n,i}|}}-\dfrac{{x}_T}{{x}_{n,i}})^2]-4(1 + \Delta_{n,i})^2 \sum_T (\dfrac{{|T|}}{{|S_{n,i}|}}-\dfrac{{x}_T}{{x}_{n,i}})^2
\end{align*}
\begin{align*}
4{\sum_T {\gamma}_o'\tilde{Z}_T -\Er^{DAX}_T \over \E(Y_{n,i})^2}&=4{1 \over \E(Y_{n,i})}[{\beta \tilde{x}_{n,i} \over \E(Y_{n,i})}\sum_T {\tilde{x}_T^2 \over \tilde{x}_{n,i}^2} +{\alpha \widetilde{|S_{n,i}|}\over \E(Y_{n,i})}\sum_T {\widetilde{|T|}^2  \over \widetilde{|S_{n,i}|}^2}-{\alpha \widetilde{|S_{n,i}|}\over \E(Y_{n,i})}\sum_T(\dfrac{\widetilde{|T|}}{\widetilde{|S_{n,i}|}}-\dfrac{\tilde{x}_T}{\tilde{x}_{n,i}})^2]\\
&-4({\alpha \widetilde{|S_{n,i}|}\over \E(Y_{n,i})})^2 \sum_T (\dfrac{\widetilde{|T|}}{\widetilde{|S_{n,i}|}}-\dfrac{\tilde{x}_T}{\tilde{x}_{n,i}})^2\\
&=2{1 \over \E(Y_{n,i})}[(1 + \Delta_{n,i})\sum_T {{x}_T^2 \over {x}_{n,i}^2} +(1 - \Delta_{n,i})\sum_T {|T|^2  \over |S_{n,i}|^2}-\\
&-(1 - \Delta_{n,i})\sum_T(\dfrac{{|T|}}{{|S_{n,i}|}}-\dfrac{{x}_T}{{x}_{n,i}})^2]-4(1 - \Delta_{n,i})^2 \sum_T (\dfrac{{|T|}}{{|S_{n,i}|}}-\dfrac{{x}_T}{{x}_{n,i}})^2
\end{align*}

If $(\dfrac{\widetilde{|T|}}{\widetilde{|S_{n,i}|}}-\dfrac{\tilde{x}_T}{\tilde{x}_{n,i}})=0$ then the regression is less accurate than areal weighting and dasymetric asymptotically. If this difference increases, the difference between the regression and the other two methods gets smaller and then the regression method can do better than the other two methods. Indeed, for example when $\dfrac{x_T}{|T|}=\dfrac{\beta x_{S_{n,i}}-\alpha|S_{n,i}|}{2 \beta |S|}$, this yields  $T, x_T$ satisfy $(\dfrac{\widetilde{|T|}}{\widetilde{|S_{n,i}|}}-\dfrac{\tilde{x}_T}{\tilde{x}_{n,i}}) \not= 0$, we have
$$\dfrac{(\gamma_o'\tilde{Z}_T)^2}{\gamma_o'\tilde{Z}_{n,i}}-{1 \over \gamma_o'\tilde{Z}_{n,i}}\beta^2 \tilde{x}_{n,i}^2(\dfrac{\widetilde{|T|}}{\widetilde{|S_{n,i}|}}-\dfrac{\tilde{x}_T}{\tilde{x}_{n,i}})^2=0$$
therefore,
$${\gamma}_o'\tilde{Z}_T -\Er^{DAW}_T=-\beta^2 \tilde{x}_{n,i}^2(\dfrac{\widetilde{|T|}}{\widetilde{|S_{n,i}|}}-\dfrac{\tilde{x}_T}{\tilde{x}_{n,i}})^2<0$$
Choosing $\eta, \eta_1$ to be sufficient small, the regression predictor is asymptotically better than the areal weighting interpolation predictor. A similar result for the case of the dasymetric predictor can be proved similarly.

We therefore proved that none of the considered three methods is always dominant.

\subsubsection{Proof of Lemma \ref{LeScRCom}}
Assume $T \in S_{n,i}$, the difference between the predictors of scaled regression and composite predictor is given by
\begin{align*}
\hat{Y}^{ScR}_T-\hat{Y}^{C}_T&=\dfrac{\hat{\gamma}_n'\tilde{Z}_T }{\hat{\gamma}_n'\tilde{Z}_{n,i} }Y_{n,i}-\dfrac{{\gamma}_o'\tilde{Z}_T }{\tilde{Z}_{n,i} {\gamma}_o}Y_{n,i}\\
&=(\hat{\gamma}_n'-{\gamma}_o')\dfrac{\tilde{\gamma}_n' \tilde{Z}_{n,i} \tilde{Z}_T -\tilde{\gamma}_n' \tilde{Z}_T \tilde{Z}_{n,i} }{(\tilde{Z}_{n,i} \tilde{\gamma}_n)^2}Y_{n,i}\\
&=(\hat{\gamma}_n'-{\gamma}_o')F_n^{-T/2}F_n^{T/2}Y_{n,i}\dfrac{\tilde{\gamma}_n' \tilde{Z}_{n,i} \tilde{Z}_T -\tilde{\gamma}_n' \tilde{Z}_T \tilde{Z}_{n,i} }{(\tilde{Z}_{n,i} \tilde{\gamma}_n)^2}
\end{align*}
where $\tilde{\gamma}_n $ belongs to the segment of $\hat{\gamma}_n$ and ${\gamma}_o$.

From Theorem \ref{ConsitencyNormality}, property \eqref{app}, conditions (C1), (C2), we have
\begin{align*}
F_n^{T/2}(\hat{\gamma}_n-{\gamma}_o) &\to_d \N(0,I)\\
F_n^{-T/2}\dfrac{Y_{n,i}-\tilde{Z}_{S_{n,i}} {\gamma}_o }{\sqrt{\tilde{Z}_{S_{n,i}} {\gamma}_o }}&\to_d 0\\
\dfrac{\tilde{\gamma}_n' \tilde{Z}_{n,i} \tilde{Z}_T -\tilde{\gamma}_n' \tilde{Z}_T \tilde{Z}_{n,i} }{(\tilde{Z}_{n,i} \tilde{\gamma}_n)^2}& \text{ bounded,}
\end{align*}
 In other words,
 $$\hat{Y}^{ScR}_T-\hat{Y}^{C}_T \to_p 0$$
 \subsubsection{Proof of Theorem \ref{TheScR}}

  Because of the boundedness of $\tilde{Z}_{n,i}$, upper boundedness of $\tilde{Z}_T$, there exists
 $$M=sup_{ \tilde{Z}_T, \tilde{Z}_{n,i}, {\gamma} \in B(\gamma_o, 1)}||\dfrac{\tilde{Z}_{n,i}\gamma_n \tilde{Z}_T -\tilde{Z}_T \gamma_n \tilde{Z}_{n,i} }{(\tilde{Z}_{n,i} \gamma_n)^2}||\E(Y_{n,i}^2)$$
 where $B(\gamma_o, 1)=\{\gamma: ||\gamma_o - \gamma ||< 1\}$. Since $\hat{\gamma}_n-{\gamma}_o \to_p 0$, the sequence $\hat{\gamma}_n, n=1,2,...$ is bounded, therefore  for any $\varepsilon>0$, when $n$ is large enough

  $$sup_{ \tilde{Z}_T, \tilde{Z}_{n,i}, \tilde{\gamma} \in segment(\gamma_o, \hat{\gamma}_n)}||\dfrac{\tilde{Z}_{n,i}\tilde{\gamma}_n \tilde{Z}_T -\tilde{Z}_T \tilde{\gamma}_n \tilde{Z}_{n,i}}{(\tilde{Z}_{n,i} \tilde{\gamma}_n)^2}||\E(Y_{n,i}^2)\1_{||\hat{\gamma}_n-{\gamma}_o||< \varepsilon}<M$$

For any $\eta >0$, there is an $\varepsilon >0$ s.t.

  \begin{align*}
  \E(\hat{Y}^{ScR}_T-\hat{Y}^{C}_T)^2\1_{||\hat{\gamma}_n-{\gamma}_o||< \varepsilon}&=\E(||\dfrac{\tilde{Z}_{n,i}\tilde{\gamma}_n \tilde{Z}_T -\tilde{Z}_T \tilde{\gamma}_n \tilde{Z}_{n,i}}{(\tilde{Z}_{n,i} \tilde{\gamma}_n)^2}||^2||Y_{n,i}||^2||(\hat{\gamma}_n-{\gamma}_o)||^2\1_{||\hat{\gamma}_n-{\gamma}_o||< \varepsilon})<M^2 \varepsilon^2 < \eta
  \end{align*}

  Evaluating the error on the set $\{||\hat{\gamma}_n-{\gamma}_o||< \varepsilon\}$, we have
  \begin{align*}
  \E(\hat{Y}^{ScR}_T-Y_T)^2\1_{||\hat{\gamma}_n-{\gamma}_o||< \varepsilon}&=\E(\hat{Y}^{ScR}_T-\hat{Y}^{C}_T)^2\1_{||\hat{\gamma}_n-{\gamma}_o||< \varepsilon}+\E(\hat{Y}^{C}_T-Y_T)^2\1_{||\hat{\gamma}_n-{\gamma}_o||< \varepsilon}\\
  &-2\E(\hat{Y}^{ScR}_T-\hat{Y}^{C}_T)(\hat{Y}^{C}_T-Y_T)\1_{||\hat{\gamma}_n-{\gamma}_o||< \varepsilon}
  \end{align*}

  Moreover
  \begin{align*}
  \E(\hat{Y}^{C}_T-Y_T)^2\1_{||\hat{\gamma}_n-{\gamma}_o||< \varepsilon}&\leq \E(\hat{Y}^{C}_T-Y_T)^2=Var(\hat{Y}^{C}_T-Y_T)=\tilde{Z_T}\gamma_o - {(\tilde{Z}_T \gamma_o)^2 \over \tilde{Z}_{n,i}\gamma_o}
  \end{align*}
  With the same argument as in theorem \ref{TheErrReg}, when $n$ is large enough
  \begin{align*}
  \E(\hat{Y}^{C}_T-Y_T)^2\1_{||\hat{\gamma}_n-{\gamma}_o||< \varepsilon}&= \E(\hat{Y}^{C}_T-Y_T)^2-\E(\hat{Y}^{C}_T-Y_T)^2\1_{||\hat{\gamma}_n-{\gamma}_o||\geq \varepsilon}\\
  &>\tilde{Z_T}\gamma_o - {(\tilde{Z}_T \gamma_o)^2 \over \tilde{Z}_{n,i}\gamma_o} - \eta
  \end{align*}

  Using a similar argument as above, we can prove $\forall \eta >0, \exists \varepsilon >0$ and $n$ large enough such that
  $$||\E(\hat{Y}^{ScR}_T-\hat{Y}^{C}_T)(\hat{Y}^{C}_T-Y_T)\1_{||\hat{\gamma}_n-{\gamma}_o||< \varepsilon}||<\eta $$

  In other words
  $$-3\eta +\tilde{Z_T}\gamma_o - {(\tilde{Z}_T \gamma_o)^2 \over \tilde{Z}_{n,i}\gamma_o}<\E(\hat{Y}^{ScR}_T-Y_T)^2\1_{||\hat{\gamma}_n-{\gamma}_o||< \varepsilon}< 2\eta +\tilde{Z_T}\gamma_o - {(\tilde{Z}_T \gamma_o)^2 \over \tilde{Z}_{n,i}\gamma_o}
  $$
  Note that $\Pr(||\hat{\gamma}_n-{\gamma}_o||< \varepsilon) \to 1$ as $n \to \infty$ and  the theorem holds.

\end{document}